\documentclass[aps,pra,groupedaddress,floatfix,showpacs,twocolumn]{revtex4-1}
\usepackage{amsmath}

\usepackage{graphicx}
\usepackage{bm}
\usepackage{epstopdf}
\usepackage[caption=false]{subfig}
\usepackage{microtype}
\usepackage{placeins}
\usepackage{braket}
\usepackage{xcolor}

\begin{document}

\title{Probabilistic Hysteresis in an Isolated Quantum System: The Microscopic Onset of Irreversibility from a Quantum Perspective}

\author{R.~B\"urkle$^{1}$}
\email{rbuerkle@rhrk.uni-kl.de}
\author{J.R.~Anglin$^{1}$}

\affiliation{$^{1}$\mbox{State Research Center OPTIMAS and Fachbereich Physik,} \mbox{Technische Universit\"at Kaiserslautern,} \mbox{D-67663 Kaiserslautern, Germany}}

\date{\today}

\begin{abstract}
Recently probabilistic hysteresis in isolated Hamiltonian systems of ultracold atoms has been studied in the limit of large particle numbers, where a semiclassical treatment is adequate. The origin of irreversibility in these sweep experiments, where a control parameter is slowly (adiabatically) tuned back and forth, turned out to be a passage blue back and forth across a separatrix (integrable case) or a passage in and out of a chaotic sea in phase space (chaotic case). Here we focus on the full quantum mechanical description of the integrable system and show how the semiclassical results emerge in the limit of large particle numbers. Instead of the crossing of a separatrix in phase space, where classical adiabaticity fails, the origin of irreversibility in the quantum system is a series of avoided crossings of the adiabatic energy levels: they become so close that already for modest particle numbers the change of the external parameter has to be unrealistically slow to reach the quantum adiabatic limit of perfectly reversible evolution. For a slow but finite sweep rate we find a broad regime where the quantum results agree with the semiclassical results, but only if besides the limit $N\to \infty$ an initial ensemble of states is considered, with sufficient initial energy width. For a single initial energy eigenstate we find in contrast that the backward sweep reveals strong quantum effects even for very large particle numbers.
\end{abstract}

\maketitle
  
\section{Introduction}
\subsection{Microscopic irreversibility in slow forward-and-back sweeps}
If a control parameter of a physical system is slowly tuned away from its initial value and then slowly tuned back again, a specific form of irreversibility can be observed in some systems: hysteresis. The system does not return to its initial state, despite the control parameter having the same value as it had initially. Hysteresis implies the breakdown of adiabaticity, and in \cite{dimer, trimer} it was demonstrated conversely that a breakdown of adiabaticity in a \textit{microscopic} system can lead to a microscopic form of irreversibility, \textit{probabilistic hysteresis}. While one might expect that the microscopic onset of irreversibility occurs through final states which gradually become different from the initial state, probabilistic hysteresis means a gradually rising probability that the final state differs dramatically from the initial one. 

Under classical mechanics the breakdown of adiabaticity that leads to probabilistic hysteresis cannot be avoided by a slower variation of the control parameter, but persists even in the quasi-static limit, because of topological changes in phase space energy shells \cite{dimer, trimer}. Such quasi-static irreversibility cannot occur under quantum mechanics, however, because in the quasi-static limit the quantum adiabatic theorem must apply. In fact it has been shown that the semiclassical and adiabatic limits in general do not commute, and so in particular for the cold-atom systems studied in \cite{dimer, trimer} the Gross-Pitaevskii mean field description can become invalid in the adiabatic limit \cite{Wu, Berry}. It is therefore not obvious whether---or how---the classical limit of probabilistic hysteresis emerges from the quantum description of these systems. Even apart from enhanced adiabaticity, furthermore, ultracold atoms can in principle show a range of other quantum effects. We therefore now study the full quantum mechanical description of a hysteresis experiment in an isolated system. In this paper we focus on the quantum version of the integrable system discussed in \cite{dimer}, i.e. the experimentally realizable two-site Bose-Hubbard ``dimer'', and leave the chaotic three-site system of \cite{trimer} for future work.

\subsection{Slow sweeps in the classical two-site Bose-Hubbard model}
The two-site Bose-Hubbard system offers a convenient toy model of quantum many-body physics, with competition between nonlinear interactions and kinetic energy in a minimal form. Since it can now be realized to a good approximation with ultracold atoms \cite{Oberthaler1, Oberthaler2}, but has also long been used as a model for Josephson junctions, the two-site Bose-Hubbard system has been studied extensively. Even on the specific subject of slow sweeps of the energy detuning between the two sites, which will be our control parameter, there have been several papers.

The majority of these Bose-Hubbard sweep papers have examined the problem within the mean-field approximation, which represents the evolution in terms of a two-state wave function, but with a cubically nonlinear Schr\"odinger equation for the time evolution of the two complex amplitudes $\alpha_{1,2}(t)$. Originally the problem was introduced by Wu and Niu \cite{Wu2}, and almost at the same time by Zobay and Garraway \cite{Zobay}, as a nonlinear generalization of the two-level Landau-Zener model for a true quantum two-state system. These papers began a tradition of referring to $|\alpha_{1,2}|^2$ as ``probabilities'', since in the true quantum two-state system they are the probabilities in projective measurements. The most important result of the early works \cite{Wu2} and \cite{Zobay} is that, unlike in the linear Landau-Zener model, $|\alpha_2|^2$ can evolve from zero to non-zero even for arbitrarily slow sweeps.

Subsequent literature has expanded this understanding of nonlinear Landau-Zener evolution. An intuitive explanation of the nonadiabaticity was more recently presented in \cite{Wu3}. In \cite{Liu} it was also found that, for sweeps starting from the initial ground state $\alpha_2=0$, the scaling of the final $|\alpha_2|^2$ with the sweep rate in the nonlinear problem is different from the linear case: it changes from an exponential dependence for weak nonlinearity (as in the linear case) to a power law dependence for stronger nonlinearity. Other effects of nonlinear Landau-Zener sweeps have been measured, or their measurement in experimentally realizable systems has been suggested; see for example \cite{Wu4, Khomeriki, Jona-Lasinio, Chen, Yang}. The nonlinear Landau-Zener effect also plays an important role in the adiabatic passage through a Feshbach resonance \cite{Itin, Tikhonenkov}.

While a single nonlinear Landau-Zener sweep thus exhibits several non-trivial features in the classical two-site Bose-Hubbard model, adding a second sweep which is the exact time-reverse of the first introduces a dramatic additional phenomenon: probabilistic hysteresis \cite{dimer}. Although $|\alpha_2|^2$ has been referred to as a probability, this terminology has simply been adopted into the nonlinear classical problem from the true two-state quantum problem; the classical evolution in the phase space corresponding to the c-number variables $\alpha_{1,2}(t)$ is deterministic. For single sweeps from low-energy initial states, moreover, the classical evolution does not even appear to be highly sensitive to initial conditions, so that there is no need to introduce a probabilistic description. Any single low-energy shell in the two-site Bose-Hubbard phase space evolves, under a single slow sweep, into another single energy shell. If the initial low energy can be controlled well in experiments, then shot-to-shot variations in the final energy, after a single slow sweep, will be small.

That is not necessarily true for higher-energy initial states, however. A slow Landau-Zener sweep in a two-site Bose-Hubbard system may split a higher-energy energy shell into \textit{two} quite distinct energy shells \cite{dimer}, so that if the initial energy can be well controlled, but not the precise initial location of the system within the energy shell, then in each repetition of an arbitrarily slow classical sweep experiment the final energy will randomly take one of two quite different final values. In such cases one can derive probabilities which are not merely a matter of referring to the classical $|\alpha_{1,2}|^2$ as probabilities, but which literally represent the random chances, in each single run of a classical experiment, of finding final $\alpha_{1,2}$ in different regions of phase space.

Initial states in the energy range that can evolve in this kind of truly probabilistic manner were not considered in mean-field literature before \cite{dimer}, presumably because they did not seem like natural initial states which could easily be prepared in experiments. They can be prepared from low-energy states, however, by a slow Landau-Zener sweep \cite{dimer}. The two-sweep forward-and-back cycle thus introduces the possibility of probabilistic hysteresis in the quasi-static limit even for low-energy initial states.

The mechanism of probabilistic hysteresis has been described and explained in \cite{dimer} entirely in the classical phase space of the mean-field model. The mean-field theory has been widely applied to the two-site Bose-Hubbard system because this kind of simple quantum many-body system usually attains this kind of correspondence with classical mechanics at large particle numbers. The fundamental question of irreversibility, however, warrants a closer check of quantum-classical correspondence in this problem.

\subsection{Slow sweeps in the quantum two-site system}
Although most studies of slow parameter sweeps in two-site Bose-Hubbard systems have been within mean-field theory, a few papers have gone beyond mean-field to the full quantum many body problem, in which there are many more orthogonal states than just two, but time evolution is governed by a \textit{linear} Schr\"odinger equation for their many complex amplitudes. It has been shown in \cite{Wu, Korsch} that even in the presence of inter-particle interactions the many-body Landau-Zener probability for non-adiabatic evolution goes to zero in the limit of infinitely slow sweeping. If a single infinitely slow sweep is thus always perfectly adiabatic in the quantum problem, the two successive sweeps of an infinitely slow forward-and-back cycle must also be adiabatic in the quantum problem. At infinite slowness, therefore, the quantum system can never reproduce the quasi-static probabilistic hysteresis of the corresponding classical system. This means that the classical and adiabatic limits do not commute for two-site Bose-Hubbard sweeps \cite{Wu,Korsch}.

It has also been shown, however \cite{Wu, Korsch, Trimborn3}, that for a fixed \textit{finitely} slow sweep rate, the quantum results \textit{do} converge onto the mean-field results with increasing particle number, with quantum-classical correspondence becoming close for total particle numbers of order 10. In \cite{Trimborn3} it was claimed that the adiabatic and classical limits only fail to commute in a \textit{single trajectory} mean-field approach, and that commutability is restored if it is the classical phase space flow of a \textit{classical ensemble} which is compared to the quantum evolution. So far, though, only single-sweep evolutions, from initial ground states, have been considered within the full quantum many-body theory of the two-site Bose-Hubbard system. In this paper we will therefore extend the fully quantum mechanical treatment of this slowly time-dependent two-site Bose-Hubbard problem beyond the single-sweep protocol, to include also the reverse sweep so that the control parameter is slowly changed in a cyclic manner.

\subsection{Main results of this paper}
Our mainly numerical analysis will confirm that if the sweep is infinitely slow then no hysteresis occurs in the quantum system. We will see, though, that the slowness needed to approach this quantum-reversible limit is the extremely slow time scale of macroscopic quantum tunnelling, which can easily become completely impractical even for quite modest particle numbers. For sweeps which are not that impossibly slow, but that are slow enough to be adiabatic in the classical problem, we will find that microscopic irreversibility \textit{does} still occur in the quantum two-site Bose-Hubbard system.

We will secondly find, however, that this classical-but-not-quantum adiabaticity is not sufficient to recover the classical result of probabilistic hysteresis with an ensemble of final states that is independent of sweep rate. This is because quantum interference effects \cite{Li}, which produce nothing dramatic after a single sweep, turn out to lead, after the second sweep, to final states that can oscillate rapidly \textit{as a function of sweep rate}, even when the sweep is within the wide range between classical and quantum adiabaticity. This failure to converge onto the classical form of probabilistic hysteresis persists \textit{even for very large particle number}.

This discrepancy between classical and quantum forms of probabilistic hysteresis cannot be removed by taking a classical ensemble. Our third main result will be rather the reverse: the classical form of probabilistic hysteresis, with no dependence on sweep rate within a wide range of very slow sweeps, is restored in the quantum two-site Bose-Hubbard system by taking a sufficiently broad ensemble of initial energy eigenstates. The mixture of sufficiently many initial energy eigenstates effectively washes out the quantum interference effects, and the incoherent summation of many Landau-Zener probabilities will be shown numerically to reproduce the classical hysteresis probability that was derived in \cite{dimer}.

\subsection{Organization of the paper}
The rest of the paper is organized as follows. In Sec.~II we explicitly introduce our simple quantum many-body system along with the parametric time dependence that will define our cyclic ``sweeps''. Sec.~III briefly reviews the semiclassical description obtained in \cite{dimer}. Sec.~IV discusses the quantum energy level structures in different dynamical regimes. Sec.~V then presents the results of numerical simulations of the sweep process and shows how the classical picture, with its two qualitatively different outcomes of the cyclic sweep experiment, emerges from the quantum system with increasing particle number. In Sec.~VI we focus on the Landau-Zener description of the sweep processes. To do so we apply the independent crossing approximation at every avoided crossing of energy levels, and show numerically that an incoherent summation of all the Landau-Zener probabilities becomes accurate if the initial energy width is sufficient, and conforms to the classical hysteresis probability if the particle number is high. We summarize our main results in Sec.~VII and offer an outlook toward future studies using quantum phase space formalisms to provide analytical insight into the non-trivial quantum-classical correspondence that we have confirmed here numerically. A final Appendix discusses different variants of the independent crossing approximation, including the more accurate modified form which we have used in our main text.

\section{Setup}
Our system is the two-mode (dimer) Bose-Hubbard system with attractive interaction $U<0$ and tunnelling rate $\Omega$; the two modes have a time-dependent energy offset $\Delta(t)$, which will be our control parameter. The system Hamiltonian therefore reads
\begin{equation}
\hat H = -\frac{\Omega}{2}(\hat{a}^{\dagger}_{1}\hat{a}_{2}+\hat{a}^{\dagger}_{2}\hat{a}_{1})+\frac{U}{2}(\hat{n}_{1}^{2}+\hat{n}_{2}^{2})+\frac{\Delta(t)}{2}(\hat{n}_{1}-\hat{n}_{2}), \label{eq:H}
\end{equation}
where the bosonic operators $\hat a_{1,2}^{\dagger}$ ($\hat a_{1,2}$) create (destroy) a boson in the respective mode 1 or 2 and the number operators $\hat n_{1,2}=\hat a_{1,2}^{\dagger} \hat a_{1,2}$ are defined as usual. In this paper we choose units such that $\hbar=1$ and measure $\Delta$, $U$, energy and time in units defined by $\Omega$. The total particle number operator $\hat N=\hat n_1+\hat n_2$ commutes with the Hamiltonian, so that the total particle number given by its eigenvalue $N$ is conserved. The classical limit of this Hamiltonian studied in \cite{dimer} is obtained by replacing the operators $\hat a_{1,2}$ by complex numbers $\alpha_{1,2}=\sqrt{n_{1,2}} \,e^{-i\varphi_{1,2}}$, such that $(\varphi_i,n_i)$ are canonical coordinates. The creation operators $\hat{a}_{1,2}^\dagger$ are replaced by the complex conjugates $\alpha_{1,2}^*$.

Our protocol consists of slowly sweeping the energy offset from a negative value $\Delta_I$ at $t=-T$ to the larger value $\Delta_0$ at $t=0$ (forward sweep) and then back again to $\Delta_I$ at $t=+T$ (backward sweep):
\begin{equation}
\Delta(t)= \Delta_{I}\frac{|t|}{T}+\Delta_{0}\left(1-\frac{|t|}{T}\right), \qquad \Delta_0>\Delta_I.
\label{eq:sweep}
\end{equation}
By sweeping ``slowly'' we mean $T \gg \Omega^{-1}$.  We will study the evolution of a quantum state through this cyclic sweep, and simply ask whether the system finally returns to its initial state or not. Before we turn to the full quantum description of this process we briefly review the semiclassical description obtained in \cite{dimer}.

\section{Semiclassical description}
\subsection{Classical Hamiltonian and microcanonical ensemble}
In the semiclassical description we evolve an ensemble of initial conditions under the mean-field equations of motion, obtained from the mean-field Hamiltonian
\begin{equation}
H=-\Omega \sqrt{p_0^2-p^2} \cos(q)+U p^2+ \Delta(t) p
\label{eq:classical_H}
\end{equation}
where $q=\varphi_1-\varphi_2$ and $p=(n_1-n_2)/2$ are classical canonical coordinates and $p_0=N/2$ is a constant. The reason why we choose an ensemble initially instead of single phase space points is twofold. If (\ref{eq:classical_H}) is interpreted as a classical system, then one can argue that typically one does not have fine control over the initial conditions but can only tune equilibrium parameters such as energy or temperature. If instead (\ref{eq:classical_H}) is interpreted as an approximation to the quantum system (\ref{eq:H}), then the evolution of an appropriate classical ensemble is the truncated Wigner approximation, which is known to give a better approximation of the quantum dynamics than the single trajectory mean-field approach.  For the sake of simplicity we therefore consider a microcanonical ensemble here, and ask what fraction of the initial classical ensemble returns to its narrow initial energy range at the end of the forward-and-back sweep as it had initially. This fraction defines the classical \textit{return probability}, which turns out not to be one in general,  even though {the sweep can be arbitrarily slow and} the classical evolution is of course exactly deterministic.

\subsection{Classical adiabaticity breakdown at a separatrix}
\begin{figure*}
\centering
\subfloat[Initial state]{\includegraphics[width=.32\textwidth]{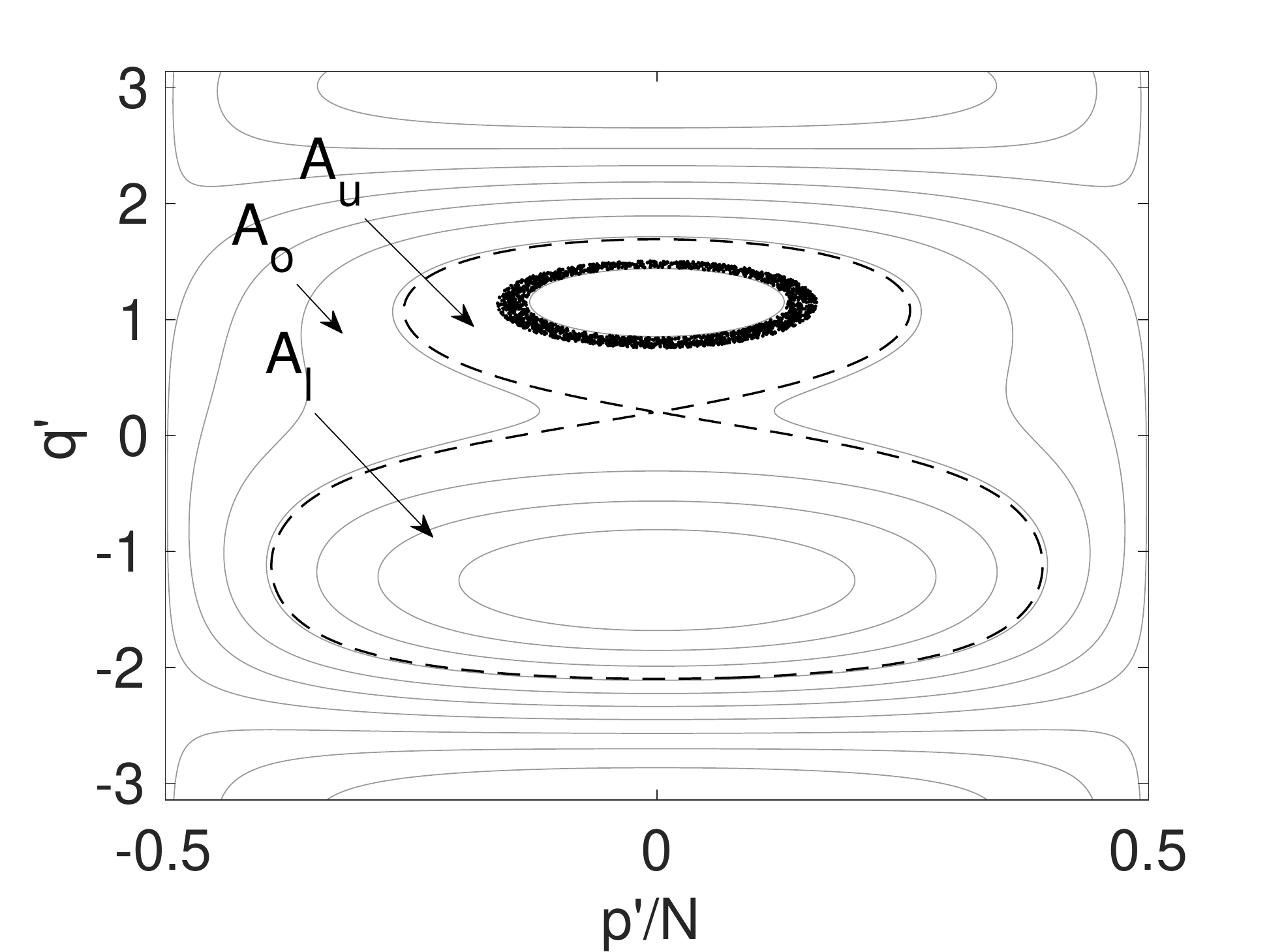}}
\subfloat[End of forward sweep]{\includegraphics[width=.32\textwidth]{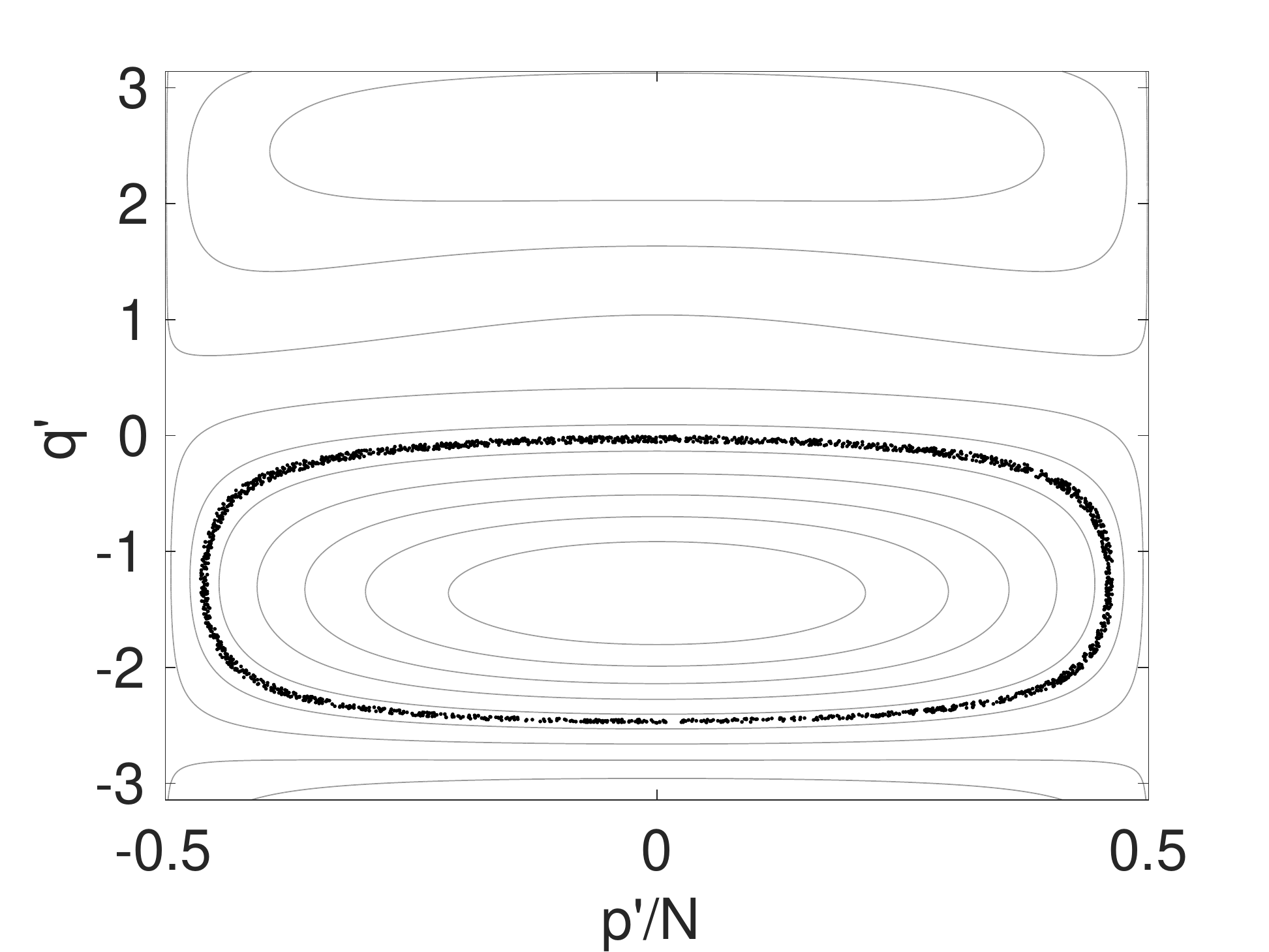}}
\subfloat[Final state]{\includegraphics[width=.32\textwidth]{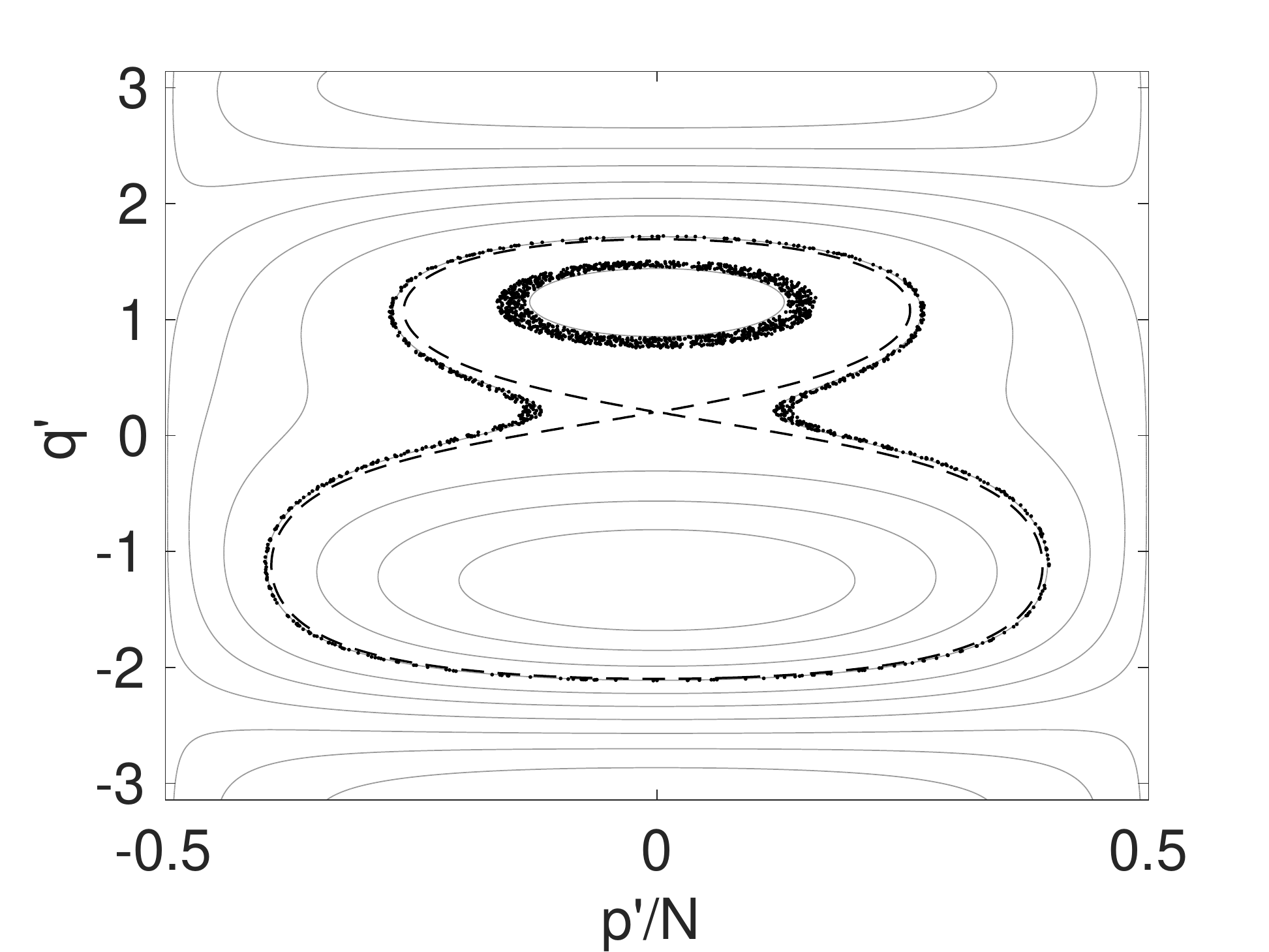}}
\caption{Evolution of a classical ensemble consisting of 2000 points (black dots) in phase space. The gray lines show adiabatic energy contours, the dashed black line is the separatrix. Because adiabaticity breaks down when the separatrix is crossed, first between (a) and (b) and then again between (b) and (c), only a finite fraction of the ensemble returns to the initial energy shell. For a clearer graphical presentation we have chosen the canonical coordinates $q'=\arctan\left(p/\left(\sqrt{p_0^2-p^2}\cos(q)\right)\right)$, $p'=-\sqrt{p_0^2-p^2}\sin(q)$ here. }
\label{fig:phase_space}
\end{figure*}

Since our sweep is assumed to be slow compared to the intrinsic time scale $\Omega^{-1}$, the classical adiabatic theorem can be applied. The action of each trajectory of the ensemble is thus an adiabatic invariant. During the forward sweep, therefore, the orbits deform in ways that keep their enclosed phase space volumes constant. During the backward sweep the same deformation happens in reverse, and so ordinarily the initial and final ensembles are expected to coincide, making the return probability always be one. This argument relies, however, on the assumption that the sweep is adiabatic during the whole sweep process. As long as the mean-field interaction $u=UN/\Omega$ is subcritical (i.e. $u>-1$ for our attractive negative $u$) this can always be fulfilled. In the supercritical case, however, there is certain range of $\Delta$ within which an unstable fixed point and a separatrix appear, as has also been demonstrated experimentally \cite{Oberthaler1}. If the initial energy is not too high, the entire ensemble is inside one lobe $A_u$ of this separatrix---see Fig.~\ref{fig:phase_space}. 
This initial separatrix lobe shrinks during the forward sweep, while the phase space area enclosed by the ensemble remains fixed adiabatically. At some point $\Delta(t)=\Delta_S$, therefore, the incompressible ensemble meets the shrinking separatrix. Since the separatrix is the orbit that runs through the unstable fixed point, this means that the orbital period of the trajectories of the ensemble diverges at this point. No matter how slow the sweep is, therefore, the condition for the adiabatic theorem can no longer be satisfied, and the actions of the trajectories change: they cross the separatrix. At this time the other separatrix lobe $A_l$ is the only phase-space region which is growing as $\Delta$ is changing; both the initial lobe $A_u$ and the region $A_o$ outside both lobes are shrinking. In accordance with Liouville's theorem, therefore, all ensemble orbits cross into $A_l$, because it is the only phase-space region which can accommodate additional orbits. After the separatrix has been crossed the orbital period becomes finite again and the evolution is again adiabatic, so that the new action is conserved during the rest of the forward sweep.

\subsection{Semiclassical probabilistic hysteresis}
During the backward sweep adiabaticity again guarantees that the evolution of the ensemble is the time-reversed evolution of the forward sweep until the separatrix is encountered again. Since at this point in the forward sweep $A_l$ was growing while both $A_u$ and $A_o$ were shrinking, now on the reverse sweep it is the other way around. The ensemble has to leave the now shrinking separatrix lobe $A_l$, but the original separatrix lobe $A_u$ as well as the outside region $A_o$ are both growing. Under incompressible Liouvillian flow, the trajectories of the ensemble must be distributed into both these growing regions. The fraction that goes to the upper separatrix lobe $A_u$---and therefore returns to the initial ensemble at the end of the sweep---is determined in the {quasi-static}  limit by the ratio of the growth rates of $A_u$ and $A_o$; this statement is known as \textit{Kruskal's theorem} (see \cite{LC} and references therein). 

Once the entire ensemble has crossed the separatrix again, its components in $A_u$ and $A_o$ evolve adiabatically in the two different regions of phase space, connected only by extremely thin threads of ensemble density that stretch between the two regions. The final state therefore consists essentially of \textit{two} ensembles with very different energies, only one of which is the initial energy. Which initial phase space points will end up in which region finally depends sensitively on initial phase space position as well as on the very slow sweep rate, but the fraction of the initial ensemble which returns to the initial energy range settles down for slow sweeps to the constant probability given by Kruskal's theorem.
In this sense there is a finite probability for each member of the initial ensemble to return to the initial energy shell (reversible evolution) or to the higher, initially unoccupied energy shell (irreversible evolution). This phenomenon was called \textit{probabilistic hysteresis} in \cite{dimer}, and the probability to return to the initial energy shell was defined as the \emph{return probability}. Note that hysteresis and irreversibility are absent in the quasi-static limit if the maximum sweep extent $\Delta_0<\Delta_S$, because then the ensemble never meets the separatrix and adiabaticity never breaks down.

In this paper we now consider how this (semi-)classical result can emerge from the full quantum description. To do so we first study the adiabatic quantum spectrum in the next Section.

\section{The quantum spectrum}
In this section we review the quantum spectrum of the Bose-Hubbard dimer and its relation to the mean-field stationary states as well as to Bohr-Sommerfeld quantization. Most of the results presented in this Section have already been presented elsewhere (see \textit{e.g.} \cite{Korsch, Graefe, Graefe2}), but are reviewed here for the reader's convenience.
\subsection{Quantum and classical adiabaticity}
\begin{figure}
\centering
\subfloat[$u=-0.5$, $N=20$]{\includegraphics[width=0.22\textwidth]{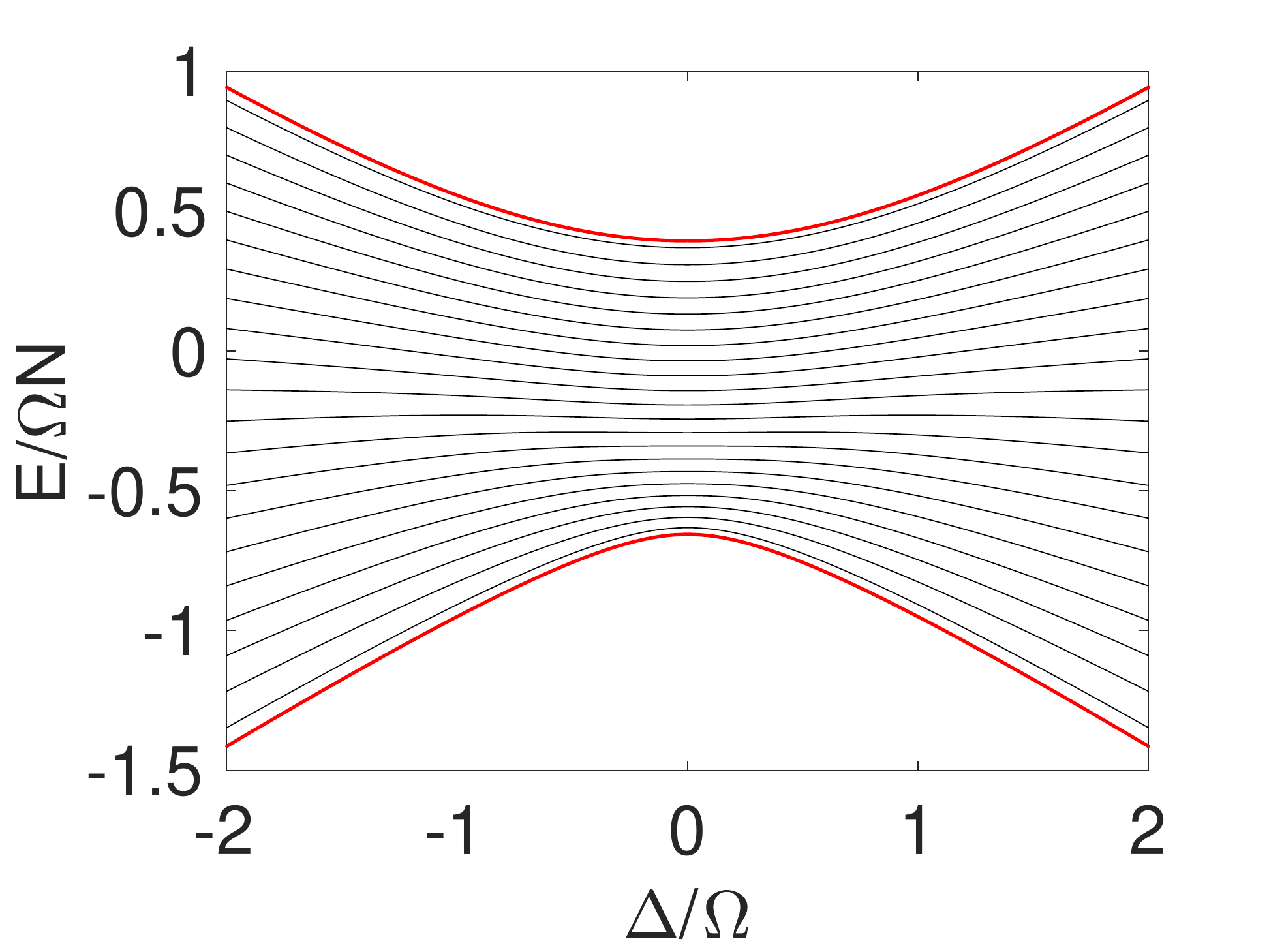}}
\subfloat[$u=-3$, $N=20$]{\includegraphics[width=0.22\textwidth]{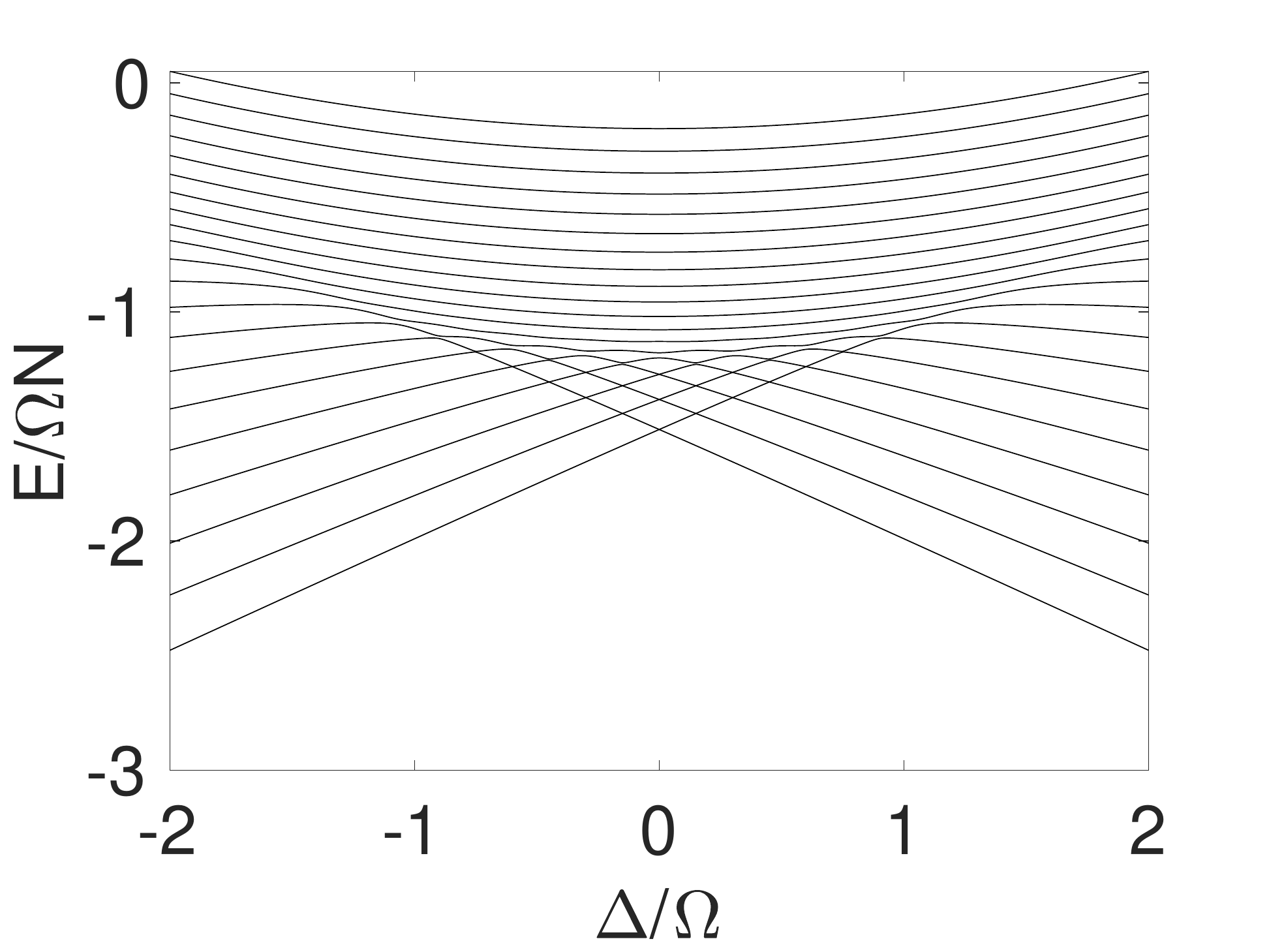}}\\
\vspace{-.3cm}
\subfloat[$u=-3$, $N=100$]{\includegraphics[width=0.22\textwidth]{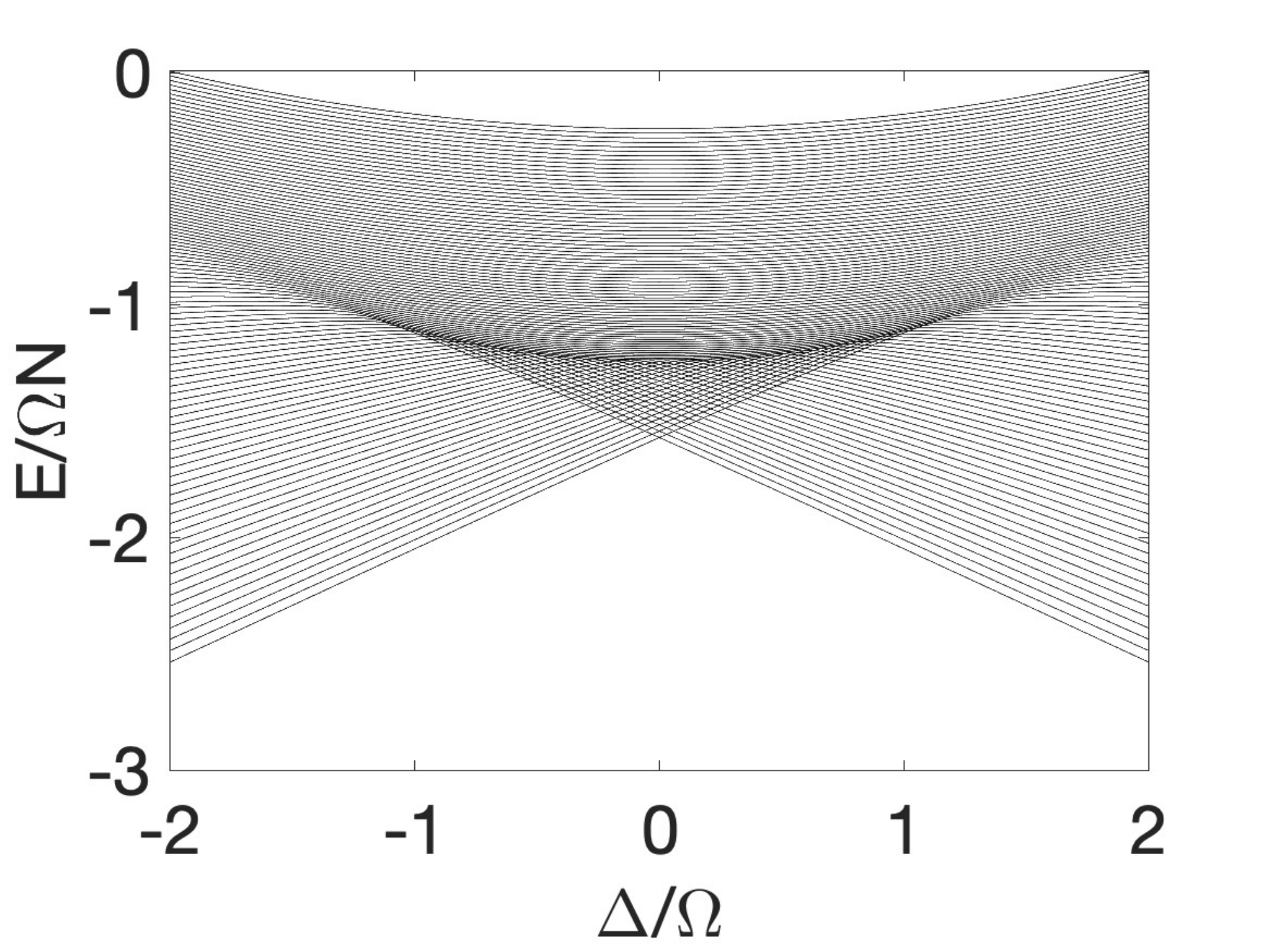}}
\subfloat[$u=-3$, $N=100$]{\includegraphics[width=0.22\textwidth]{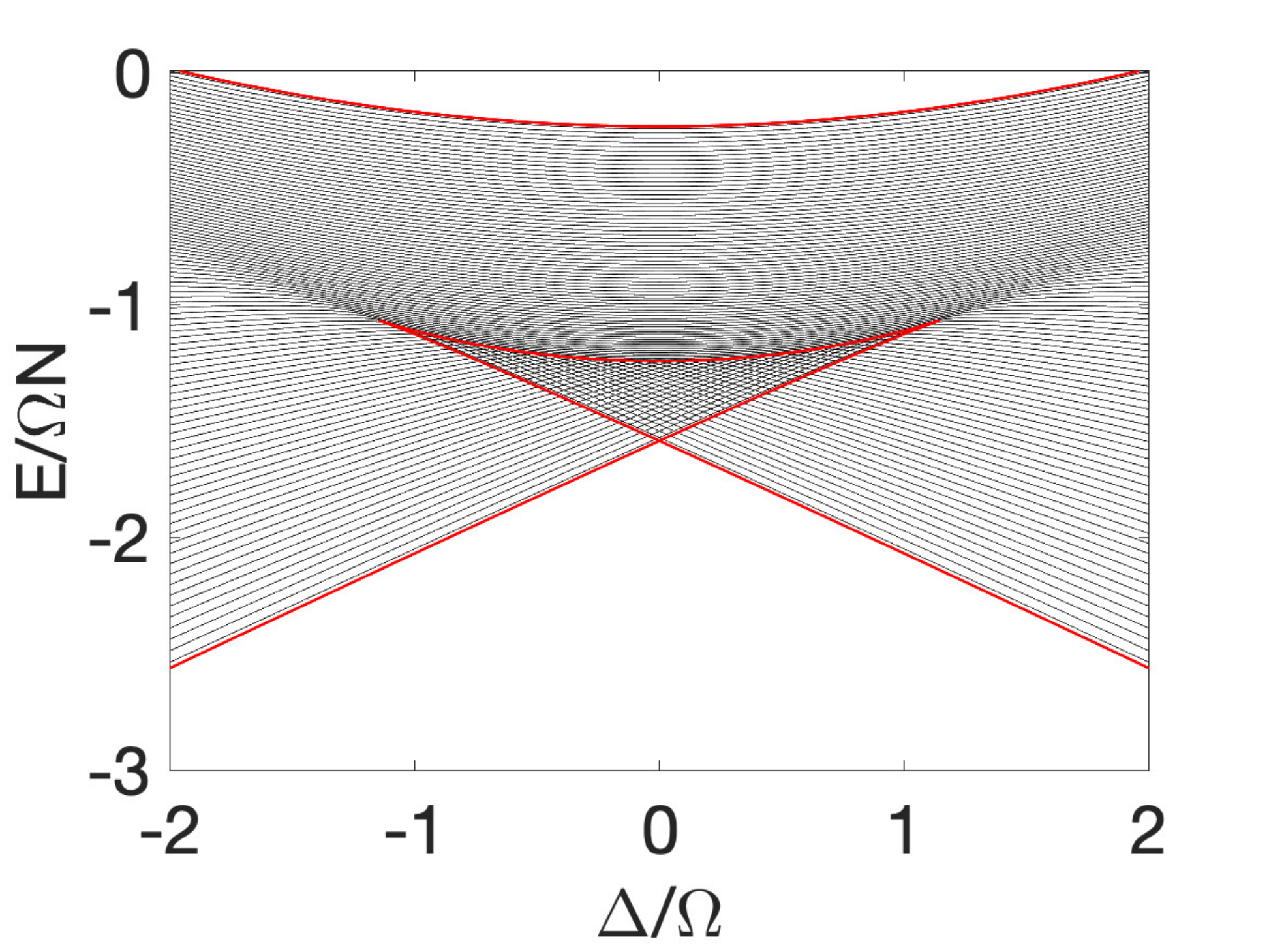}}
\caption{Eigenvalues of the Bose-Hubbard dimer in the subcritical case (a) and supercritical case (b)--(d). All level crossings are avoided. In the subcritical case (a) no extremely narrow avoided crossings are present. In the supercritical case (b), (c), however, many energy gaps become extremely small. Panel (d) shows the energies of the stationary states of the mean-field Hamiltonian in red on top of the same quantum system shown in (c). The swallowtail structure of the classical stationary states can be seen clearly in the quantum spectrum: the stable classical stationary states correspond to quantum energy eigenvalues while the unstable classical state traces an arc of avoided crossings.}
\label{fig:spectrum}
\end{figure}

The two essentially non-classical features of the quantum system both stand out in Fig.~\ref{fig:spectrum}, showing the quantized energy eigenvalues of $\hat{H}$ from (\ref{eq:H}) as functions of detuning $\Delta$. The first non-classical feature is simply that the plots are full of curving lines: energy levels are quantized. The vertical spacings between successive lines are mostly of order $\Omega$, with little dependence on $N$.

As long as the quantum energy levels remain separated from each other by order $\Omega$, the ``slow sweep'' condition $\Omega T \gg 1$ which makes the classical evolution adiabatic everywhere except near a separatrix also makes the quantum evolution adiabatic. If the system begins in an energy eigenstate, and the levels remain separated by order $\Omega$, then during a slow sweep of $\Delta$ the system will remain in the same energy eigenstate, as the energy eigenvalue slowly changes, in accordance with the quantum adiabatic theorem \cite{Landau-Lifshitz}. In the ``sub-critical'' regime with mean-field interaction strength $|u|<1$, there is never any classical separatrix because the attractive nonlinearity is too weak to ever support self-trapping, and so the classical evolution remains adiabatic throughout both forward and backward slow sweeps, giving perfect reversibility \cite{dimer}. 

\begin{figure}
\includegraphics[width=0.45\textwidth]{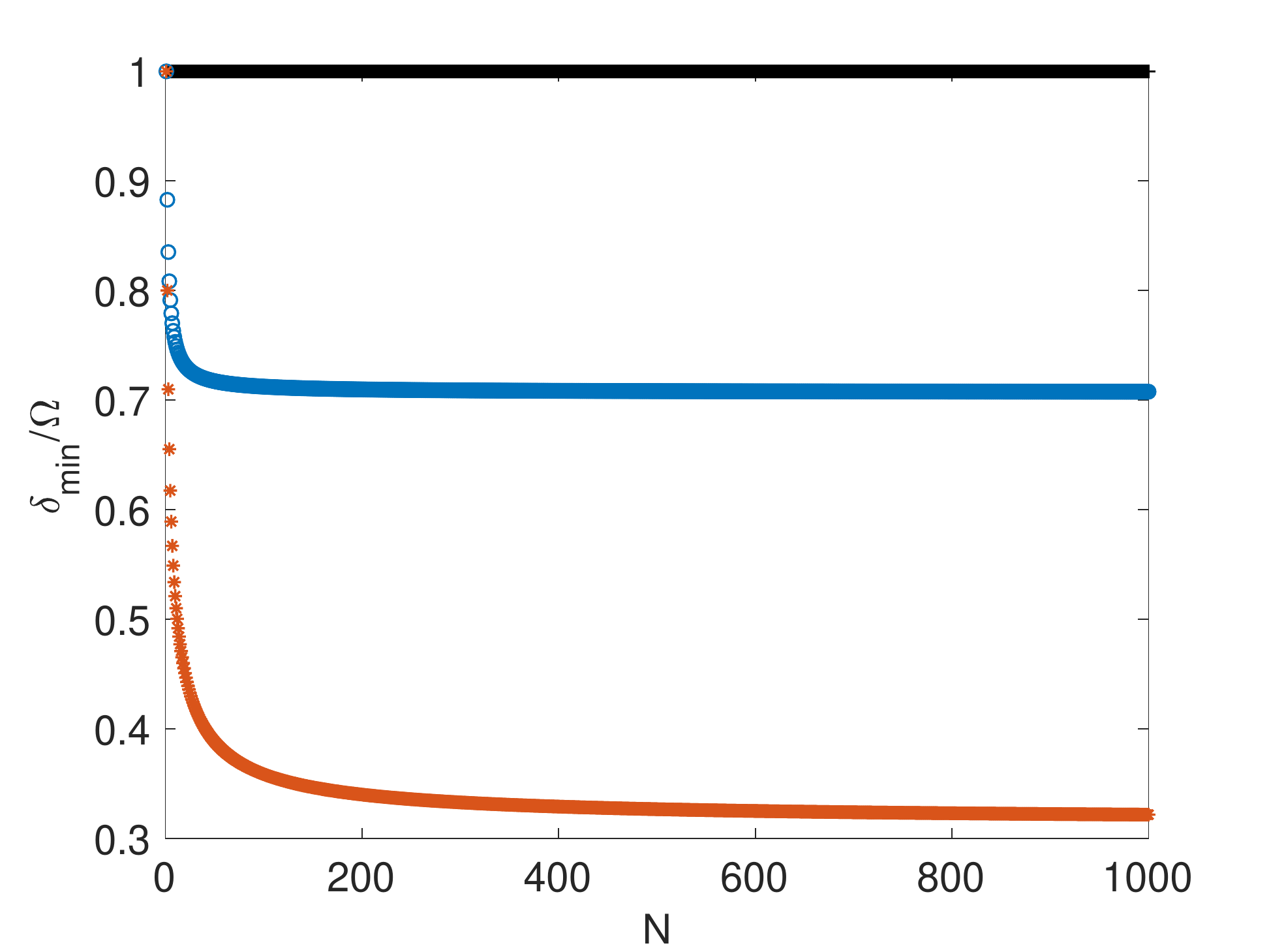}
\caption{Minimal level spacing $\delta_{\mathrm{min}}$ in the spectra for subcritical $u=0$ (top, black $+$), $u=-0.5$ (middle, blue $\circ$) and $u=-0.9$ (bottom, red $*$). The size of level spacing quickly settles to a constant finite value for increasing particle number $N$.}
\label{fig:subcritical_spacings}
\end{figure}

As Fig.~\ref{fig:spectrum}(a) indicates for $u=-0.5$ and $N=20$, the quantum energy spectrum in the sub-critical regime never has any levels approach each other more closely than order $\Omega$. The energy gaps $\delta$ between successive pairs of energy levels are all smallest at $\Delta=0$, with the smallest individual gap $\delta_{\mathrm{min}}$ being between the lowest two levels. As the particle number increases the absolute size of this minimal energy gap decreases, but it quickly settles to a finite value $\delta_{\mathrm{min}}^*/\Omega=\sqrt{1+u}$ \cite{Wu} depending only on $u$; see Fig.~\ref{fig:subcritical_spacings}.
Since all other energy gaps are even larger than $\delta_{\mathrm{min}}$, the quantum adiabatic limit at sub-critical $u$ can easily be reached for every initial state \cite{Landau-Lifshitz}, even for arbitrarily large particle numbers, and the evolution is fully reversible since the system stays in the same adiabatic level during the whole forward-and-back sweep. Quantum-classical correspondence is thus straightforward in the sub-critical regime, where probabilistic hysteresis does not occur.

\subsection{The  quantum ``separatrix''}
In the super-critical regime $u<-1$, however, things are more complicated. As Fig.~\ref{fig:spectrum}(b)-(d) show for different $N$ at $u=-3$, the second distinctly non-classical feature of the quantum system now shows up in the energy spectrum: crossings and brief close approaches (``avoided crossings'') of eigenvalue curves in the $(E,\Delta)$ plane. These crossing features are all found within the inverted triangle ``swallowtail'' region of the $(E,\Delta)$ plane, which is bounded by the energies of the classical fixed points (see Fig.~\ref{fig:spectrum}(d)). The connection between these classical energies and the quantum levels is simply Bohr-Sommerfeld quantization, which becomes highly accurate at large particle number $N$ for this classically {integrable} system. Under Bohr-Sommerfeld quantization the quantum energy levels are found by quantizing the actions of classical orbits, so that in a Bohr-Sommerfeld system quantum and classical adiabaticity generally coincide. 

\begin{figure}
\centering
\includegraphics[width=.45\textwidth]{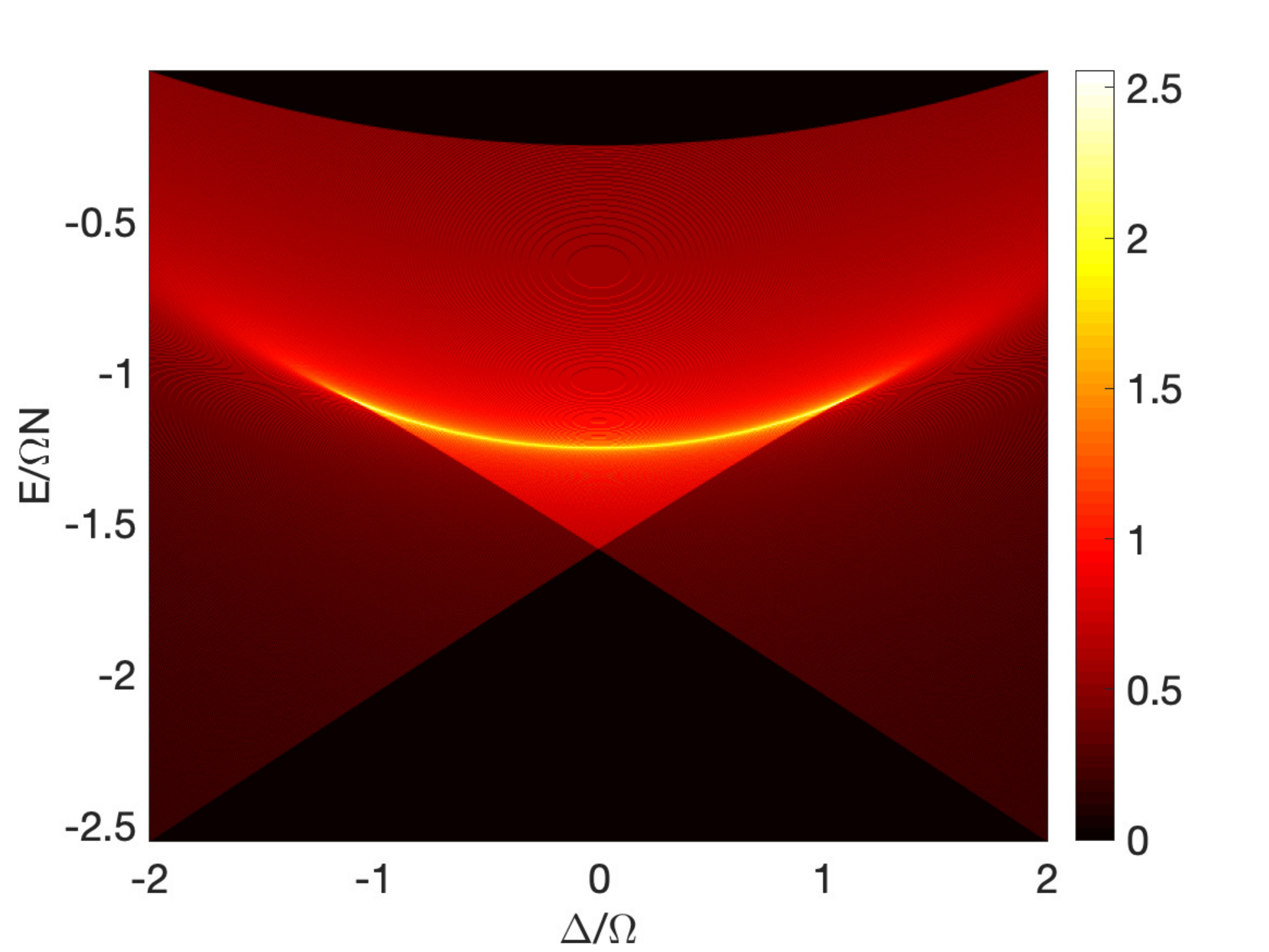}
\caption{Density of states for $N=1000$ and $u=-3$.}
\label{fig:dos}
\end{figure}

Simple geometry dictates, however, that when $u<-1$ there are crossings of Bohr-Sommerfeld energy levels within the swallowtail {which is} traced by the classical fixed points in the $(E,\Delta)$ plane, as well as a rather dense accumulation of intersecting lines along the inverted arch at the top {of} the swallowtail. We will refer to this upper border of the swallowtail feature as ``the separatrix'', even though it is a curve in the $(E,\Delta)$ plane rather than in phase space, because the particular curve which defines the upper edge of the swallowtail is precisely the energy of the classical separatrix. Bohr-Sommerfeld theory explains why the separatrix shows up as a peak in the quantum density of states, which for large $N$ becomes very  sharp, with energy levels packing closely together \cite{Aubry, Buonsante}. See Fig.~\ref{fig:dos}. 
This connection between the density of states and the classical unstable fixed point follows from the energy $E=I \omega$ in action-angle coordinates $(I,\omega=2\pi/\mathcal{T})$, where $\mathcal{T}$ is the orbital period, together with the Bohr-Sommerfeld quantization rule $I_n=2 \pi(n+1/2)\hbar$
\begin{equation}
\mathcal{T}=2\pi \frac{\mathrm dI}{\mathrm dE}=4\pi^2 \hbar \frac{\mathrm dn}{\mathrm dE}.
\end{equation}
Therefore the unstable fixed point, where the orbital period $\mathcal{T}$ diverges, corresponds to a maximum in the density of states $\mathrm d n/\mathrm dE$ \cite{Graefe, Graefe2}.

\subsection{Avoided crossings}
The top of the swallowtail does not correspond to a single Bohr-Sommerfeld level, however; nor does the high density of states around the separatrix represent many eigenvalues running parallel to each other at small separations. Instead the high density of states within this narrow arc represents a network of avoided crossings. As $N$ becomes large the network of crossings becomes dense in the $(E,\Delta)$ plane, and the minimal separations at each individual avoided crossing become narrower. One can see this pattern by looking closely at Fig.~\ref{fig:spectrum}(b), for which $N=20$; at $N=100$ the network has already become too dense to see clearly in Fig.~\ref{fig:spectrum}(c).

As a matter of fact there are no actual crossings of any energy levels at all. All of the apparent crossings in Fig.~\ref{fig:spectrum} are really avoided crossings, because a tridiagonal matrix cannot have degenerate eigenvalues \cite{Parlett}, and $\hat H$ is tridiagonal in the Fock basis. The avoidances of most of these crossings are impossible to see at larger $N$, though, because the energy gaps with which they are avoided become extremely narrow \cite{Karkuszewski, Aubry, Buonsante}. 

The two lowest-lying eigenvalues which seem to cross at $\Delta=0$, for example, are the energies of states with the majority of the attractively interacting particles localized in Bose-Hubbard site 1 or site 2, respectively. These energies avoid crossing because the unique ground and first excited states at $\Delta=0$ are both mesoscopic ``Schr\"odinger's Cat'' superpositions, with relative phase 0 or $\pi$ respectively, of these two significantly different distributions of particles. The tiny energy difference between these even and odd superposition states is due to mesoscopic quantum tunnelling; it is exponentially small in $N$. Below we will confirm numerically that the other avoided crossings within the swallowtail region are all likewise exponentially small for large $N$. 

The avoidance of Bohr-Sommerfeld crossings due to dynamical tunnelling turns out to be the crucial feature for probabilistic hysteresis in quantum systems. Outside the $(E,\Delta)$ swallowtail, quantum level spacings are all of order $\Omega$ and quantum adiabaticity is assured as long as $\Omega T\gg 1$, as we always assume. All evolution outside the swallowtail region is thus essentially trivial; everything significant happens because of the avoided crossings within the swallowtail. 

\subsection{Non-classical quantum adiabaticity}
To illustrate the implications of these very narrowly avoided crossings, suppose that the system's initial state is the ground state at large negative $\Delta$. In a slow upward sweep of $\Delta$, therefore, the system follows the lowest energy level up to the avoided crossing of the first two levels at $\Delta=0$.  If the sweep is really perfectly slow such that $T \to \infty$ (quasi-static limit), the system stays in the adiabatic ground state through the narrowly avoided crossing and continues to follow the lowest energy level. When the sweep returns the avoided crossing is encountered again, and by the same argument the system inevitably ends up back in the initial ground state at $t=+T$. In this extreme case of infinite slowness our sweep process is completely reversible, just as in the subcritical case. 

\begin{figure*}
\centering
\subfloat[]{\includegraphics[width=.32\textwidth]{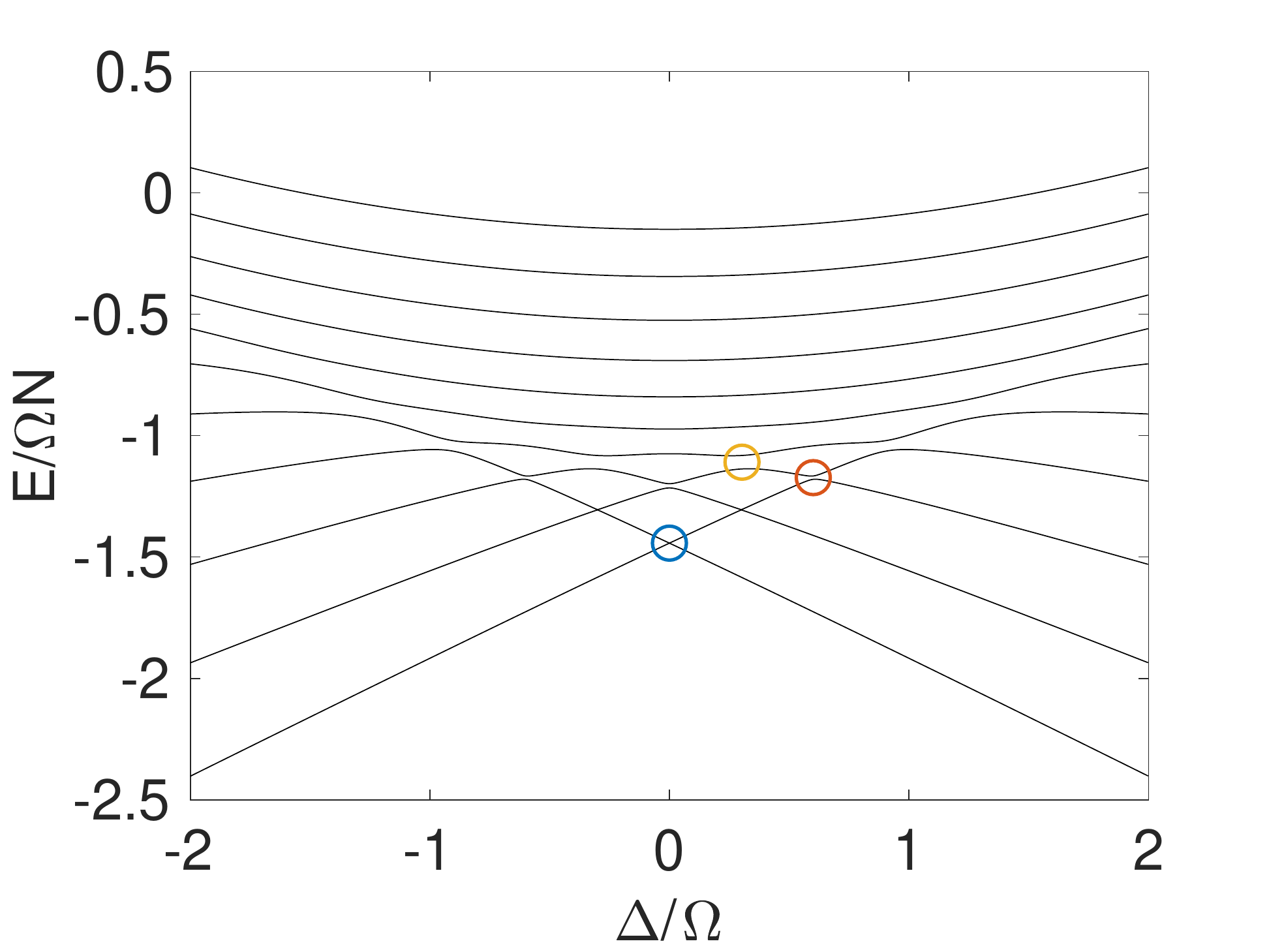}}
\subfloat[]{\includegraphics[width=.32\textwidth]{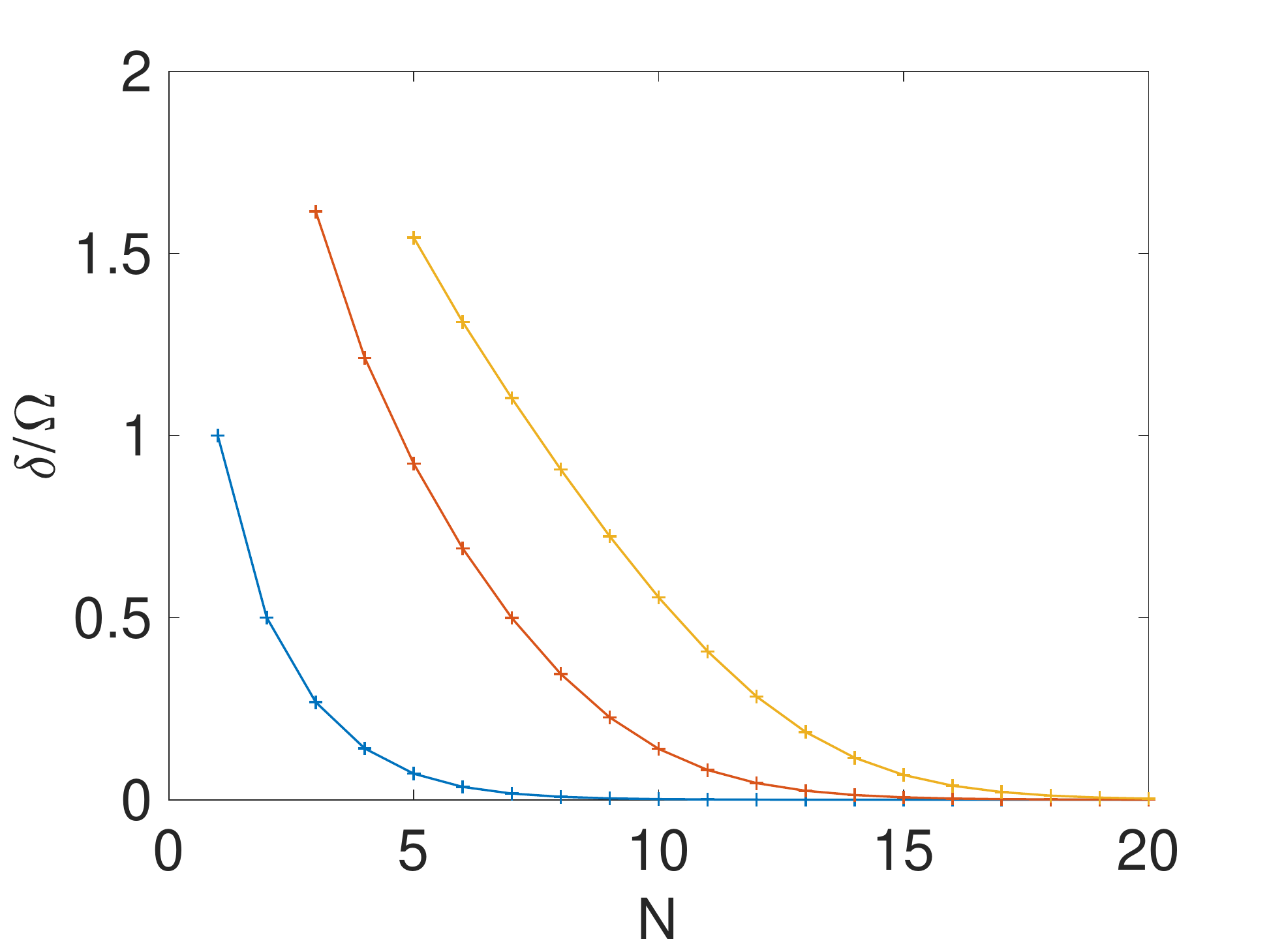}}
\subfloat[]{\includegraphics[width=0.32\textwidth]{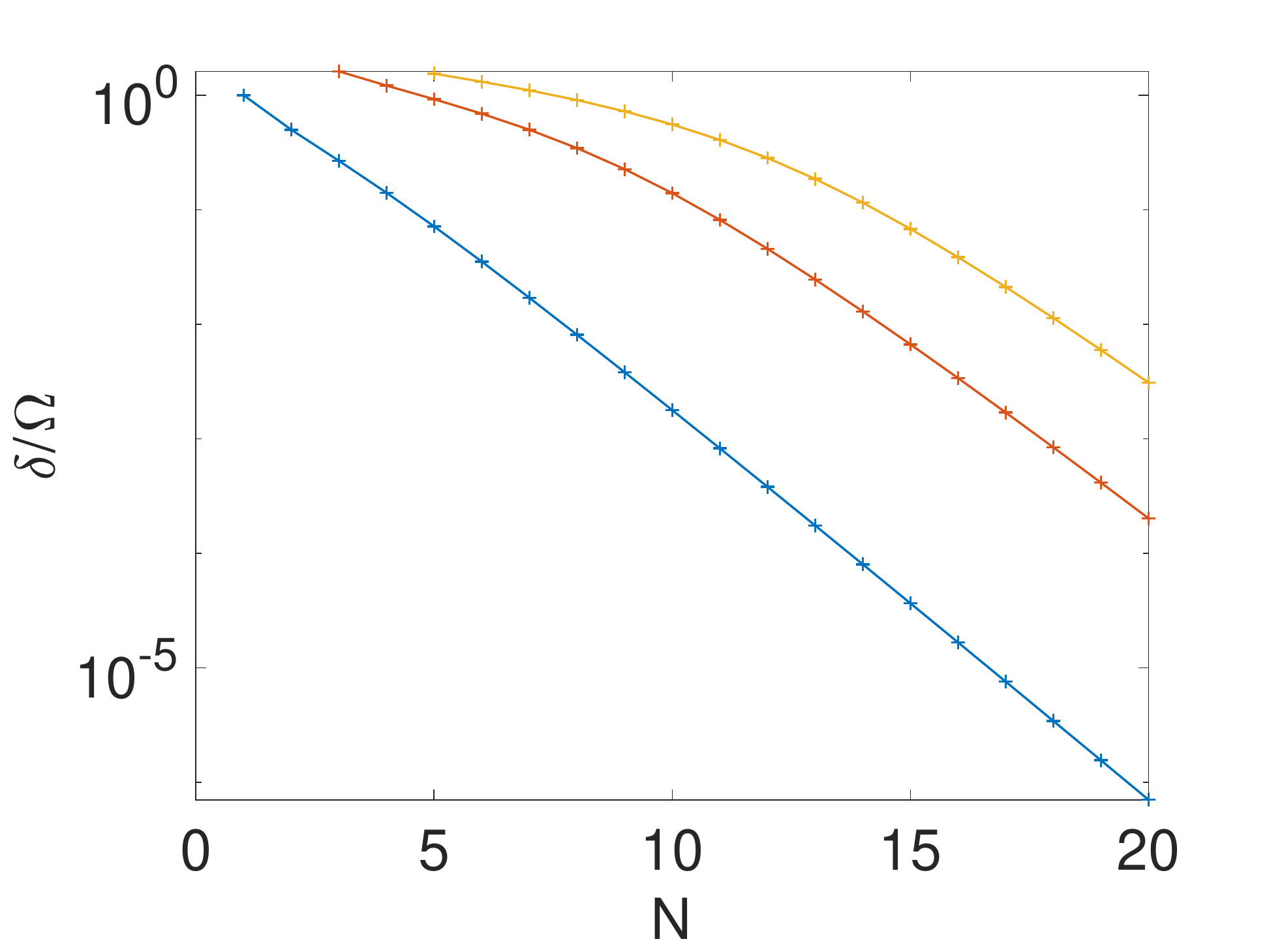}}
\caption{Size of the energy gap $\delta$ at three different avoided crossings for $u=-3$ in dependence of the particle number. The energy gaps do not approach a finite value as in the subcritical case but are exponentially small in $N$ (for $N$ not too small) as we show in (b),(c) with linear or logarithmic $y$-axis, respectively. As examples we have considered the avoided crossings marked in (a): the avoided crossing of the first two levels (blue),  the rightmost avoided crossing of the third and fourth levels, and the second avoided crossing between the fourth and fifth levels from the right.  The lines in (b) and (c) simply connect the numerical points, as guides to the eye.}
\label{fig:first_spacing}
\end{figure*}

In any real experiment, however, the sweep time $2T$ is necessarily finite, and the validity of the adiabatic approximation depends on the size of the energy gaps $\delta$ at the avoided crossings. Fig.~\ref{fig:first_spacing} shows these energy gaps at three avoided crossings as a function of the particle number $N$ as an example. 
We find that with increasing particle number $N$ the energy gaps of the narrowly avoided crossings quickly become exponentially small in $N$ and do not settle to a constant non-zero value. In fact this is true for all avoided crossings within the swallowtail structure. With rising particle number, therefore, it quickly becomes increasingly difficult to fulfill the condition of the quantum adiabatic theorem. Already for moderate particle numbers and slow but finite sweep rates there is a significant probability to follow the diabatic path at the first avoided crossing, making a transition to the second adiabatic energy level. As the sweep continues, more avoided crossings are encountered and the same reasoning can be applied. Whether or not the system ends up in its initial state when the forward-and-back sweep is completed now becomes a non-trivial question. In the next Sections we will determine this non-trivial evolution numerically for a representative range of particle numbers.

\section{Regimes of particle number}
Throughout the rest of this paper we will take $u=-3$ as a representative supercritical case; all other cases $u<-1$ are essentially similar. The range of particle numbers in the supercritical case can be divided for $u=-3$ into three qualitatively different regimes:
\begin{itemize}
\item[(i)] quantum adiabatic regime ($N\lesssim20$)
\item[(ii)] irreversible quantum regime ($20 \lesssim N \lesssim 50$)
\item[(iii)] classical correspondence regime ($N \gtrsim 50$)
\end{itemize}
In the following we will discuss these regimes briefly and show the results obtained by numerically solving the Schr\"odinger equation. Except in regime (i) we will choose the same sweep rate for all simulations. 

\subsection{Regime (i): quantum adiabatic}
\begin{figure*}
\centering
\subfloat[Adiabatic spectrum]{\includegraphics[width=.32\textwidth]{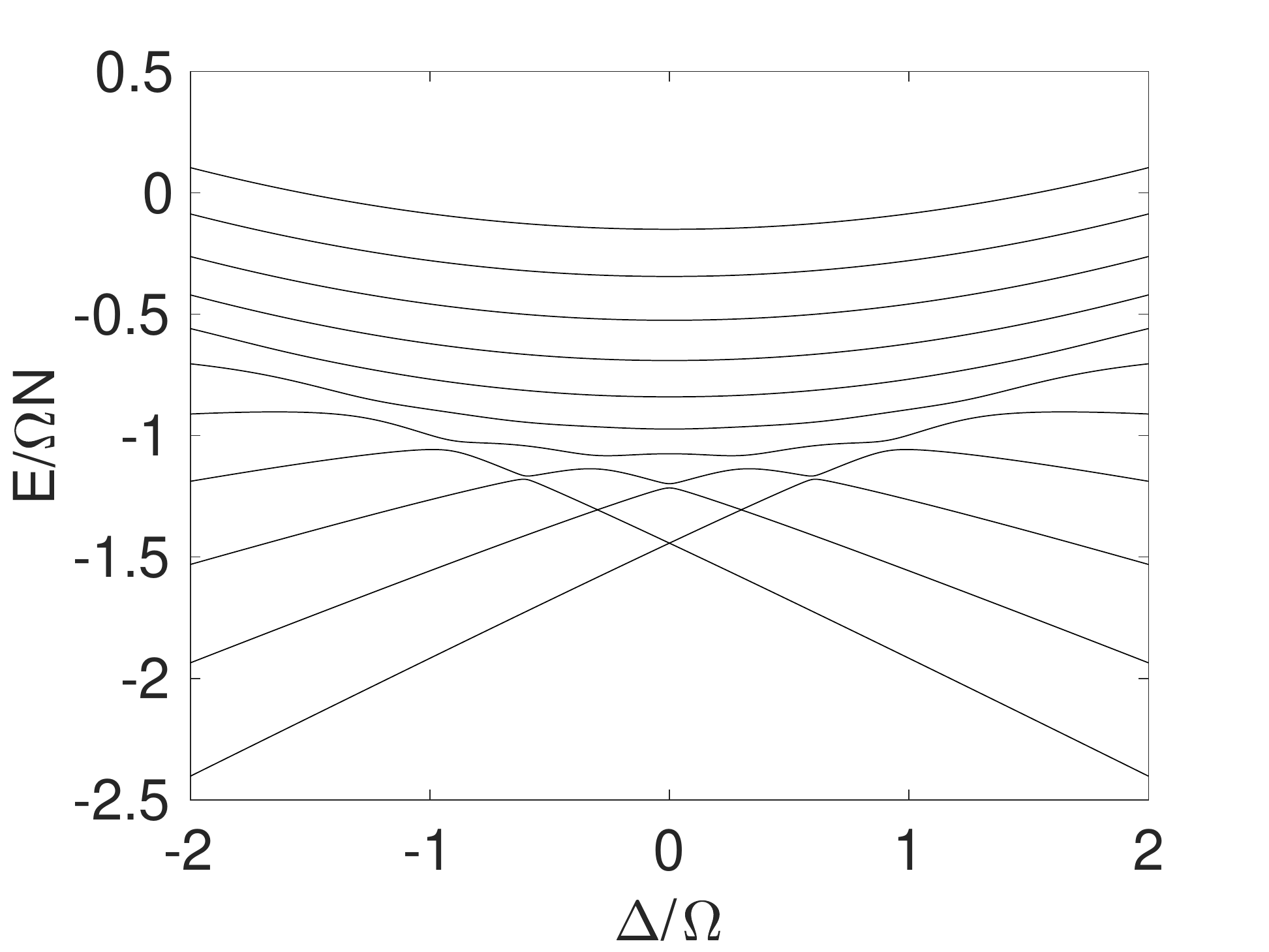}}
\subfloat[Forward sweep]{\includegraphics[width=.32\textwidth]{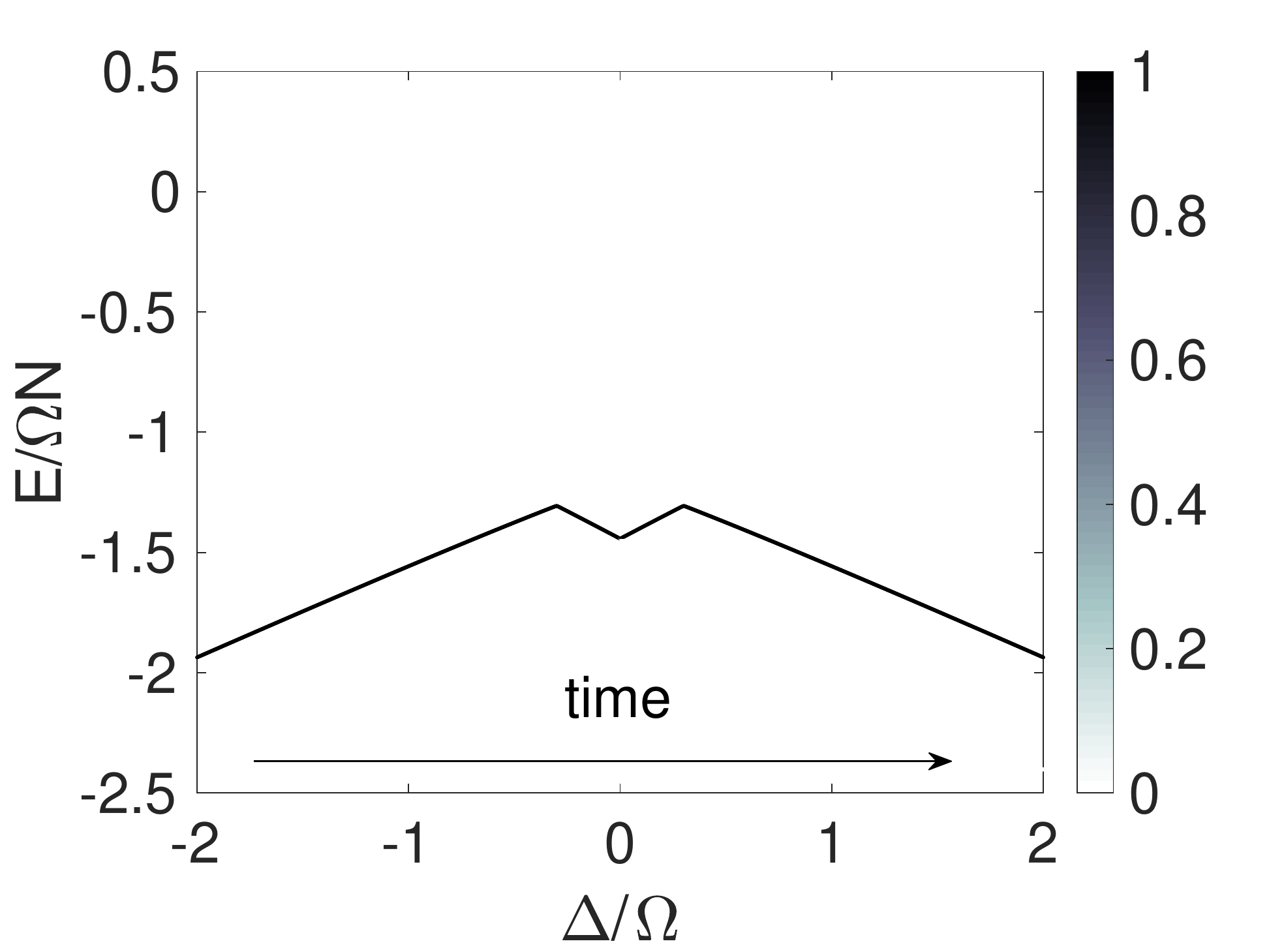}}
\subfloat[Backward sweep]{\includegraphics[width=.32\textwidth]{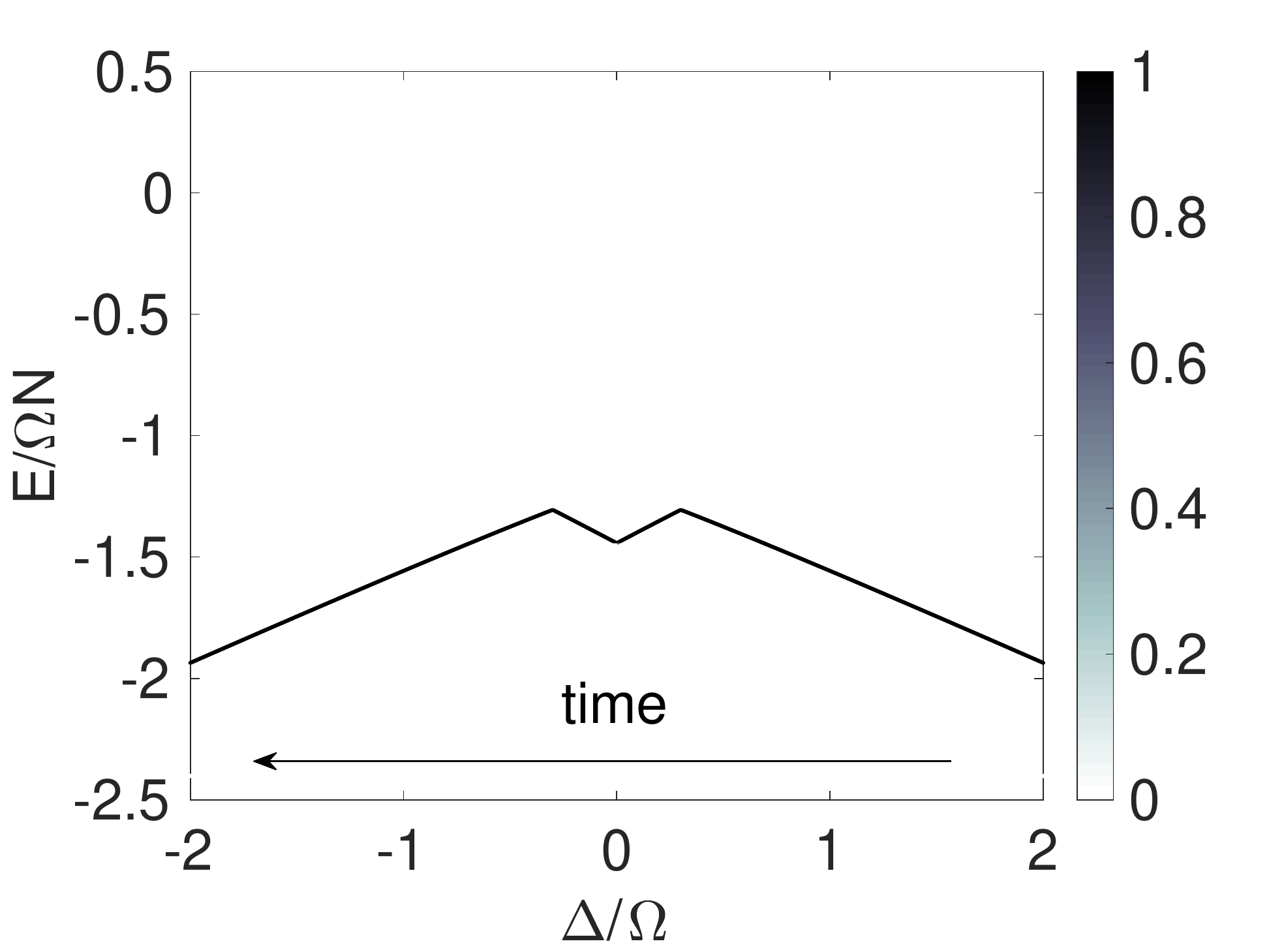}}
\caption{Quantum adiabatic regime: Panel (a) shows the adiabatic eigenenergies for $N=10$. Panels (b) and (c) show the probability to find the system in the adiabatic eigenstate with energy $E$ during the forward and backward sweep with $T=10^8\Omega^{-1}$, starting from the second-lowest initial energy eigenstate (as an example). In this case the system always stays in the same adiabatic eigenstate and so the return probability is one. Note that in panel (b) time runs from left to right (forward sweep) and in panel (c) from right to left (backward sweep).}
\label{fig:N=10}
\end{figure*}

In the quantum adiabatic regime \textit{all} the level spacings are large enough that the quantum adiabatic limit can realistically be reached, i.e. the sweep can be made so slow that the system always stays in the adiabatic state in which it started. Accordingly the return probability is always one and does not depend on the sweep rate, as long as it is slow enough. We emphasize that due to the exponential smallness of most of the energy gaps this regime can be hard to reach even for small particle numbers, so that we have to choose a much slower sweep rate than in the other regimes. An example for $N=10$ and $T=10^8 \Omega^{-1}$ is shown in Fig.~\ref{fig:N=10}.
For $N=20$ and the same parameters, the total sweep time $2T$ already has to be on the order of $10^{15} \Omega^{-1}$ to obtain a fully reversible evolution, which would be around 30 years if $\Omega$ were in the experimentally typical MHz regime, or even 30000 years for $\Omega$ in the experimentally feasible kHz regime. (These estimates of minimum sweep time are based on the Landau-Zener formula that we will discuss in Sec.~\ref{sec:LZ}, assuming generously that ``fully reversible'' evolution means a diabatic transition probability of around 1\% or less.)

\subsection{Regime (ii): irreversible quantum}
\begin{figure*}
\centering
\subfloat[Relevant part of the adiabatic spectrum]{\includegraphics[width=.32\textwidth]{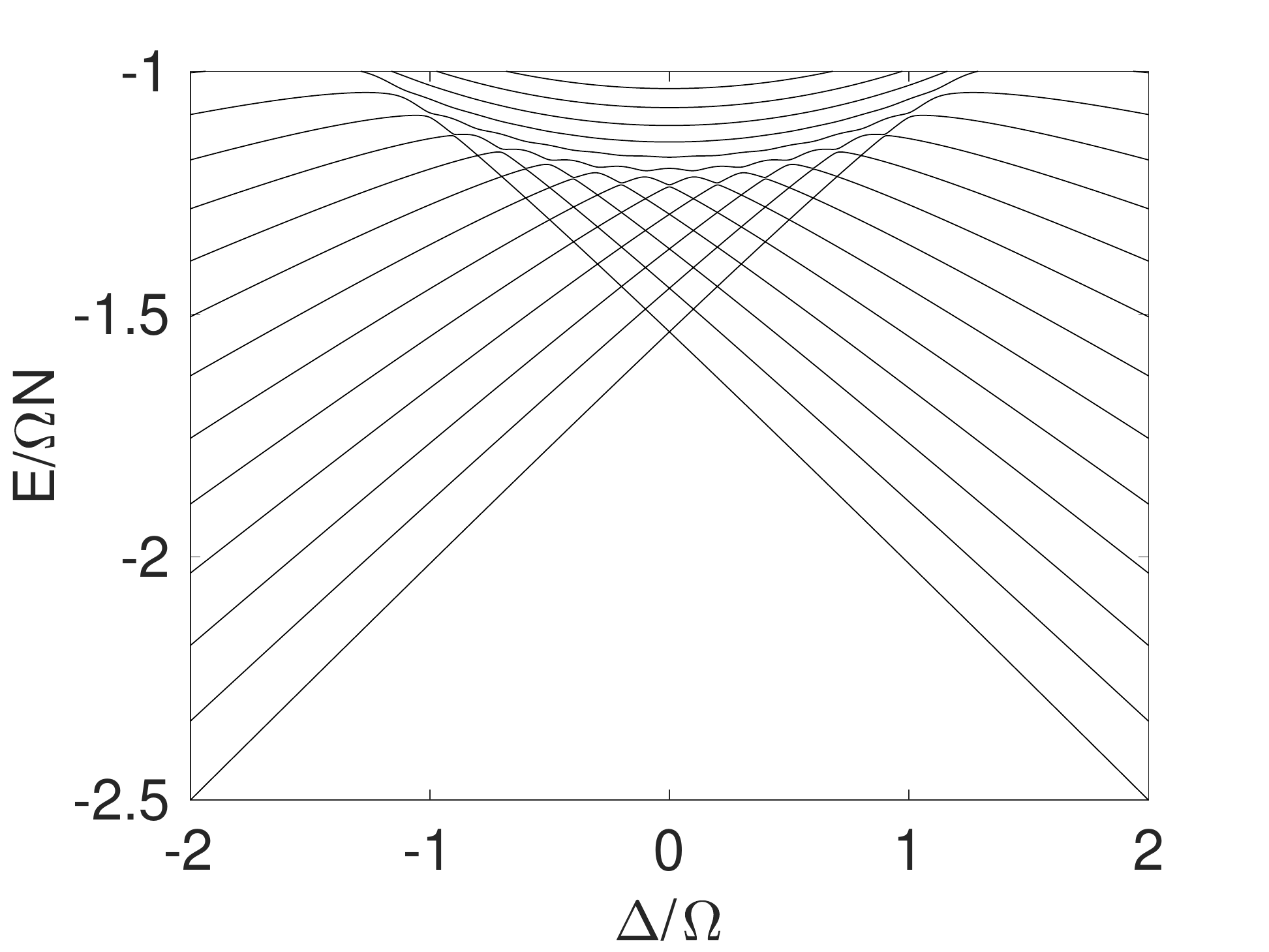}}
\subfloat[Forward sweep]{\includegraphics[width=.32\textwidth]{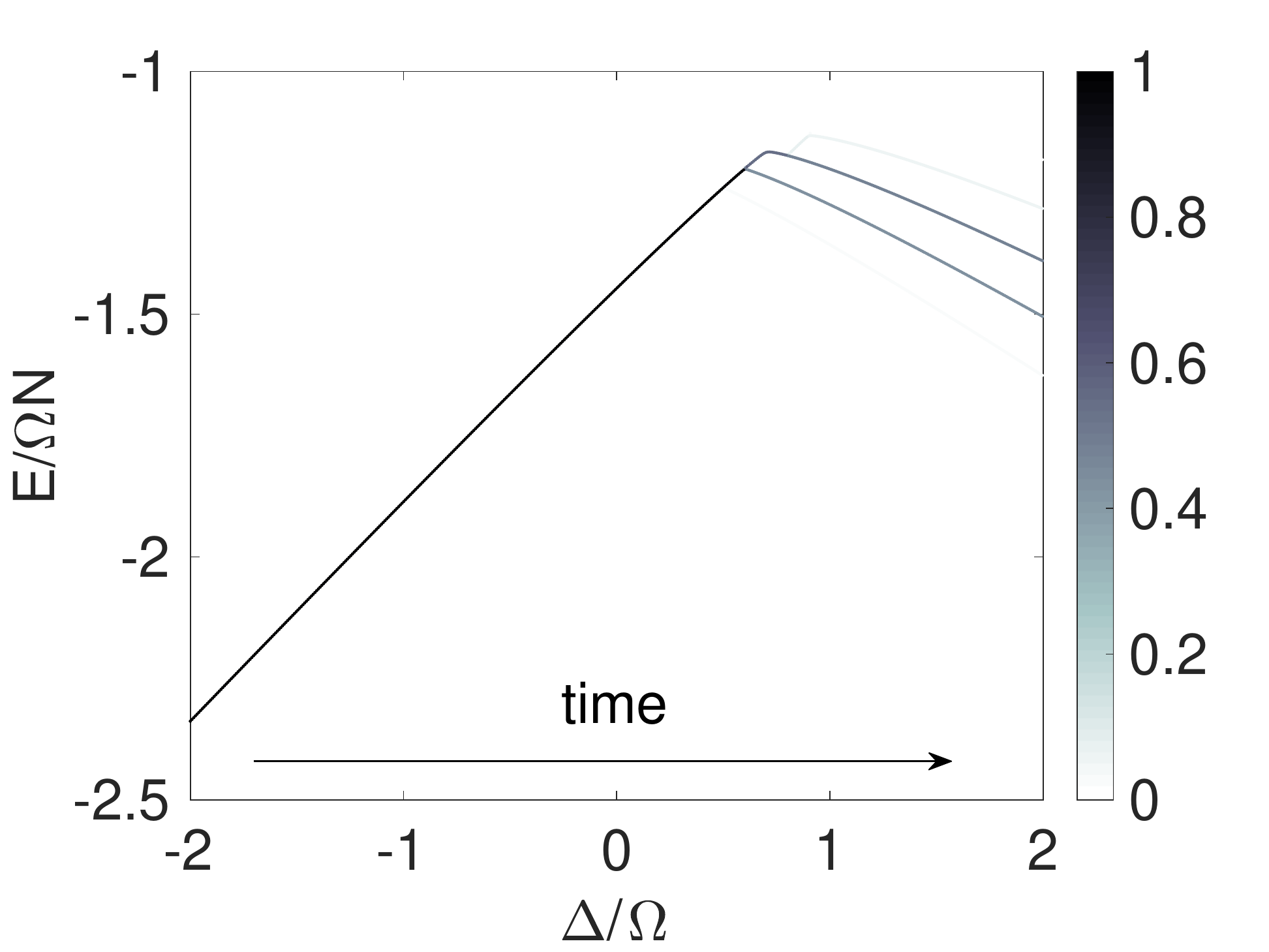}}
\subfloat[Backward sweep]{\includegraphics[width=.32\textwidth]{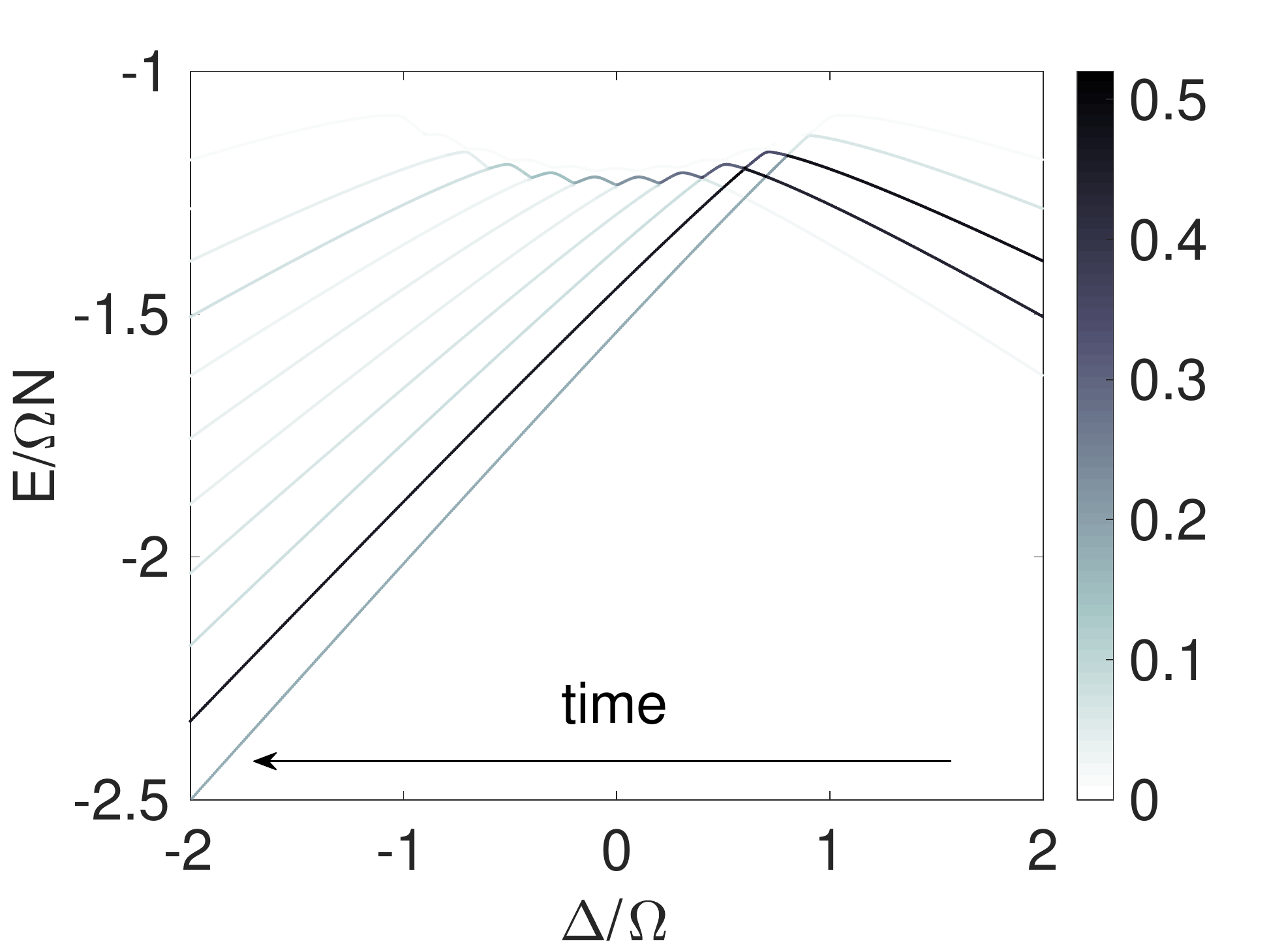}}
\caption{Irreversible quantum regime: Panel (a) shows the relevant adiabatic eigenenergies for $N=30$. Panels (b) and (c) show the probability to find the system in the adiabatic eigenstate with energy $E$ during the forward and backward sweep with $T=5000\Omega^{-1}$. Unlike in Fig.~\ref{fig:N=10} the probability spreads over many adiabatic eigenstates. To stay in the same adiabatic eigenstate and obtain a fully reversible evolution as in Fig.~\ref{fig:N=10} the sweep time would have to be on the order of $10^{22} \Omega^{-1}$, an unattainable requirement for practically any realizable $\Omega$. Here again we have shown evolution from the second-lowest initial eigenstate as an example.} 
\label{fig:N=30}
\end{figure*}

As the particle number is increased it becomes rapidly more and more difficult to reach the quantum adiabatic limit. At some point we may say that the evolution has become irreversible for all practical purposes, since sweep times of many thousand years (!) are clearly impractical. When this form of irreversibility sets in the initial state is in general not recovered with unit probability after any realistically feasible slow sweep; see Fig.~\ref{fig:N=30}.
After the forward sweep in regime (ii), therefore, the system is not in a single adiabatic eigenstate any more, but in a coherent superposition of adiabatic eigenstates. In the backward sweep the probability spreads further so that the initial state is recovered with only about $50\%$ probability. The rest of the final probability is spread over a range of energies around the initial energy. Notice, however, that the final probability distribution is quite different from what we found semiclassically. In the semiclassical case we had two well-separated final energy shells, one coinciding with the initial energy shell and one having much higher energy. In Fig.~\ref{fig:N=30}, however, this clear separation of final energies is not present. Although we therefore find irreversibility in regime (ii), it is significantly different from the irreversibility we found in the semiclassical system.

\subsection{Regime (iii): classical correspondence}
\begin{figure*}
\centering
\subfloat[Relevant part of the adiabatic spectrum]{\includegraphics[width=.32\textwidth]{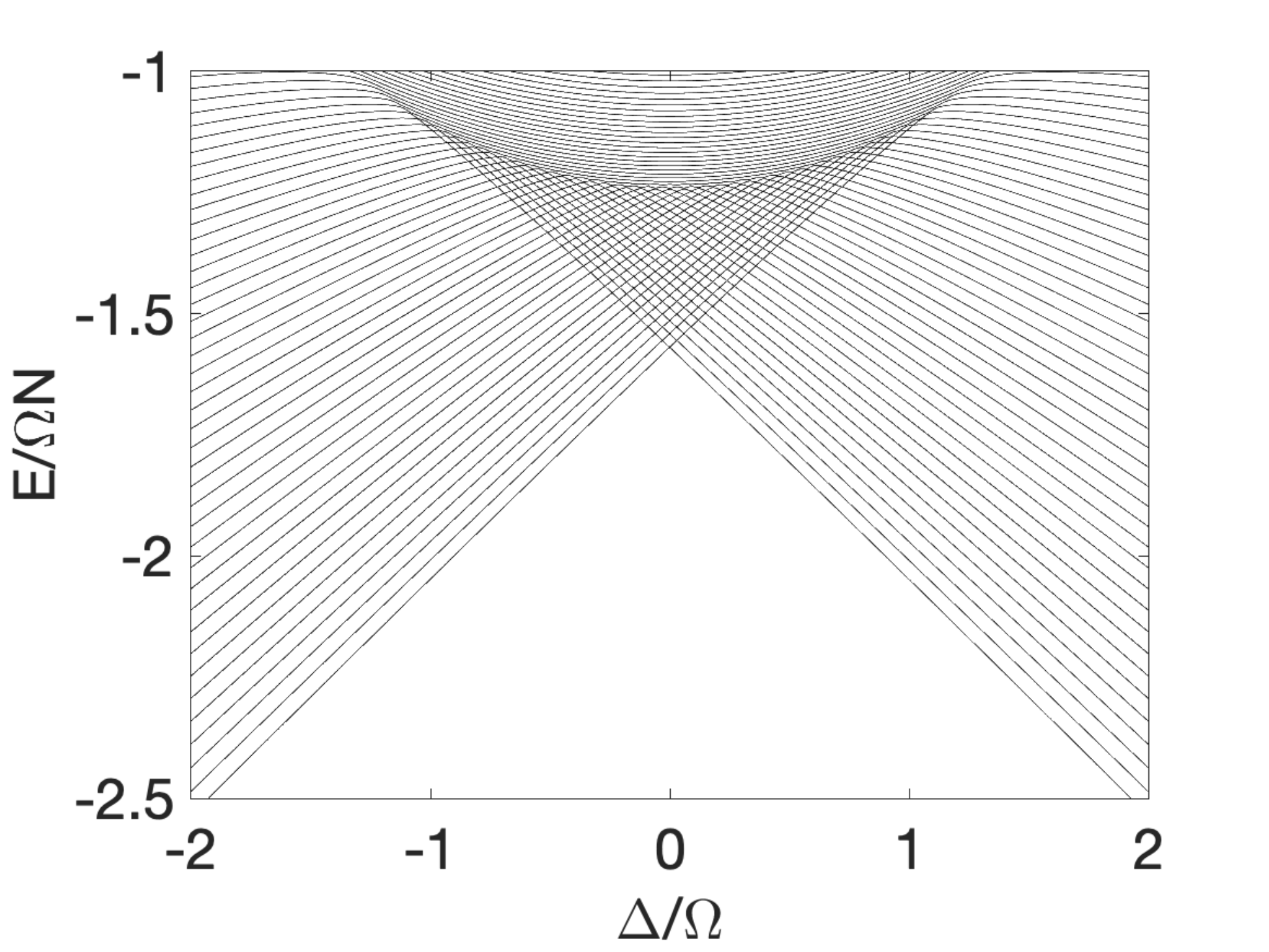}}
\subfloat[Forward sweep]{\includegraphics[width=.32\textwidth]{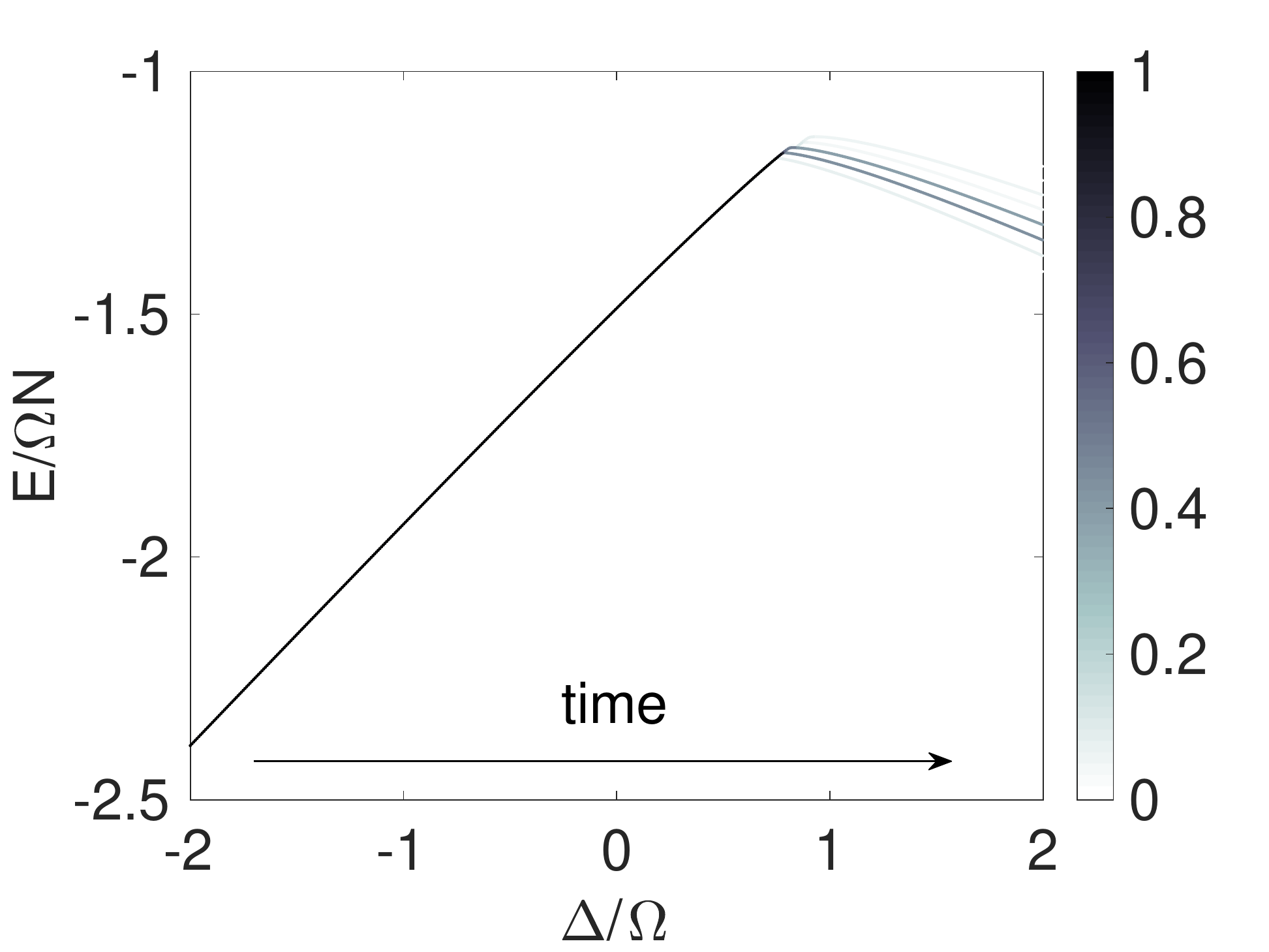}}
\subfloat[Backward sweep]{\includegraphics[width=.32\textwidth]{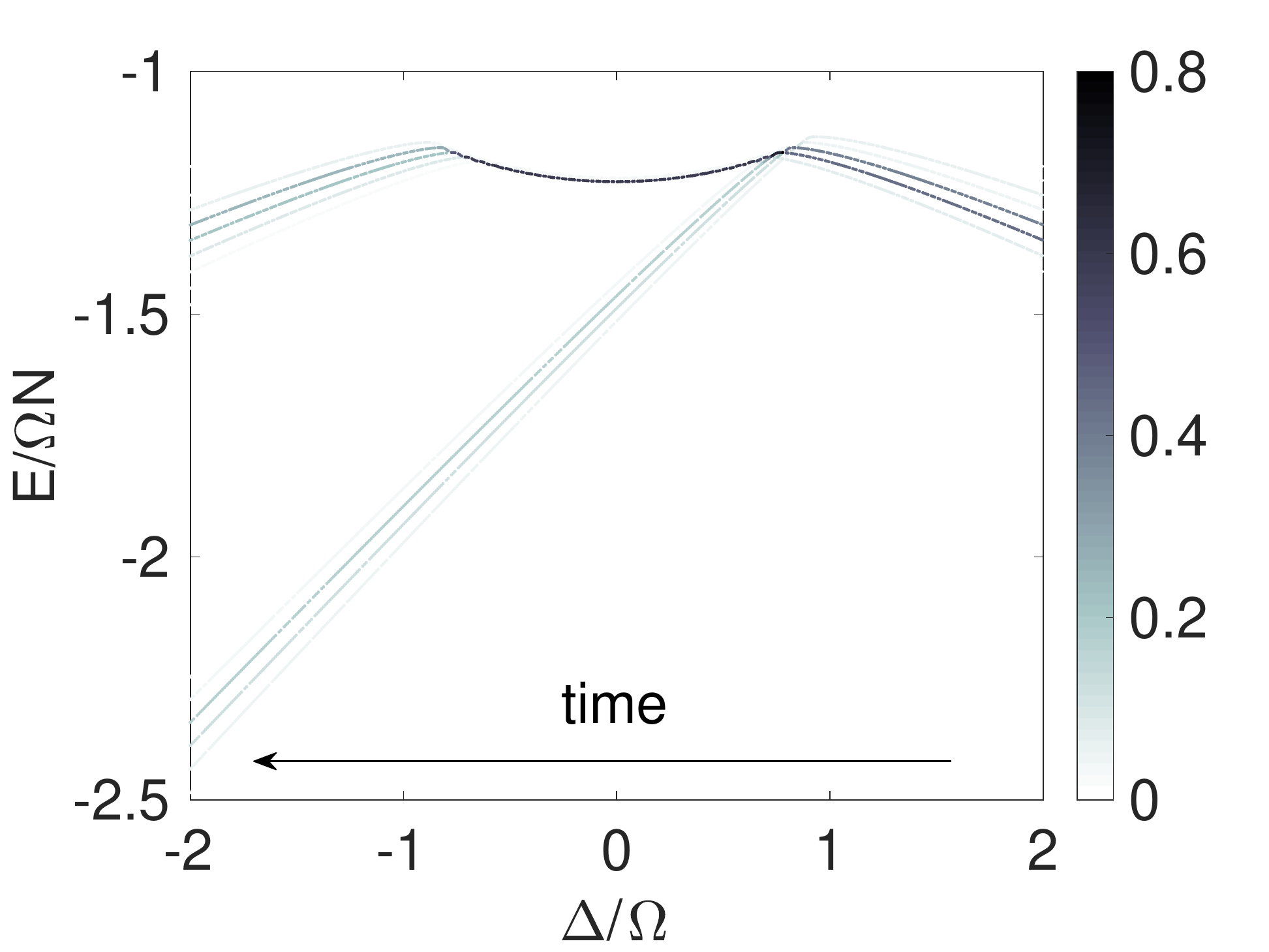}}
\caption{Classical correspondence regime: Panel (a) shows the relevant adiabatic eigenenergies for $N=100$. Panels (b) and (c) show the probability to find the system in the adiabatic eigenstate with energy $E$ during the forward and backward sweep with $T=5000\Omega^{-1}$. In panel (c) we see a clear bifurcation of the probability during the backward sweep into two branches, as in the corresponding classical problem. Here we have chosen the fourth-lowest initial eigenstate as the example because it has approximately the same energy per particle as the initial state of Fig.~\ref{fig:N=30}.}
\label{fig:N=100}
\end{figure*}

\begin{figure*}
\centering
\subfloat[Forward sweep]{\includegraphics[width=.32\textwidth]{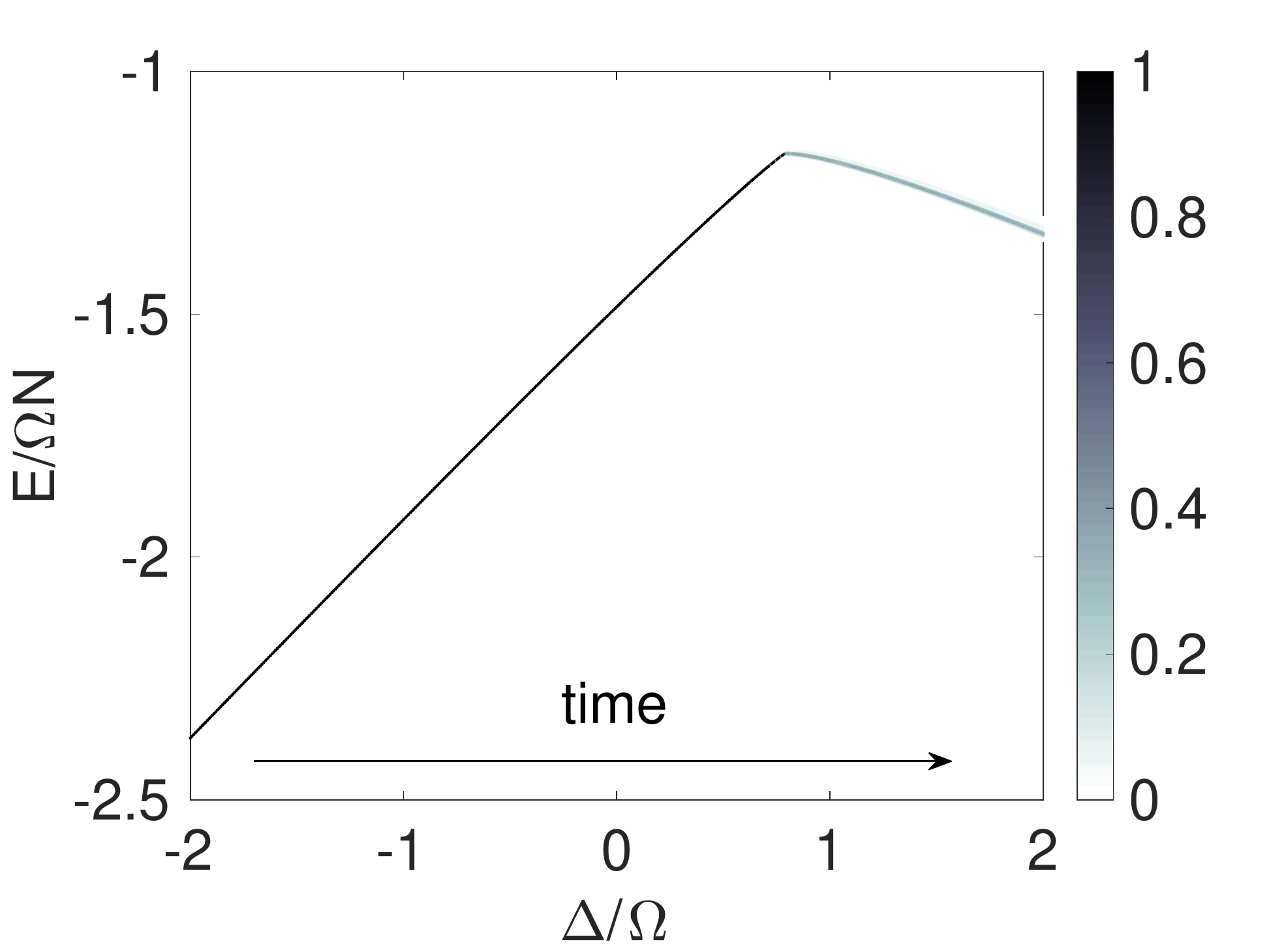}}
\subfloat[Backward sweep]{\includegraphics[width=.32\textwidth]{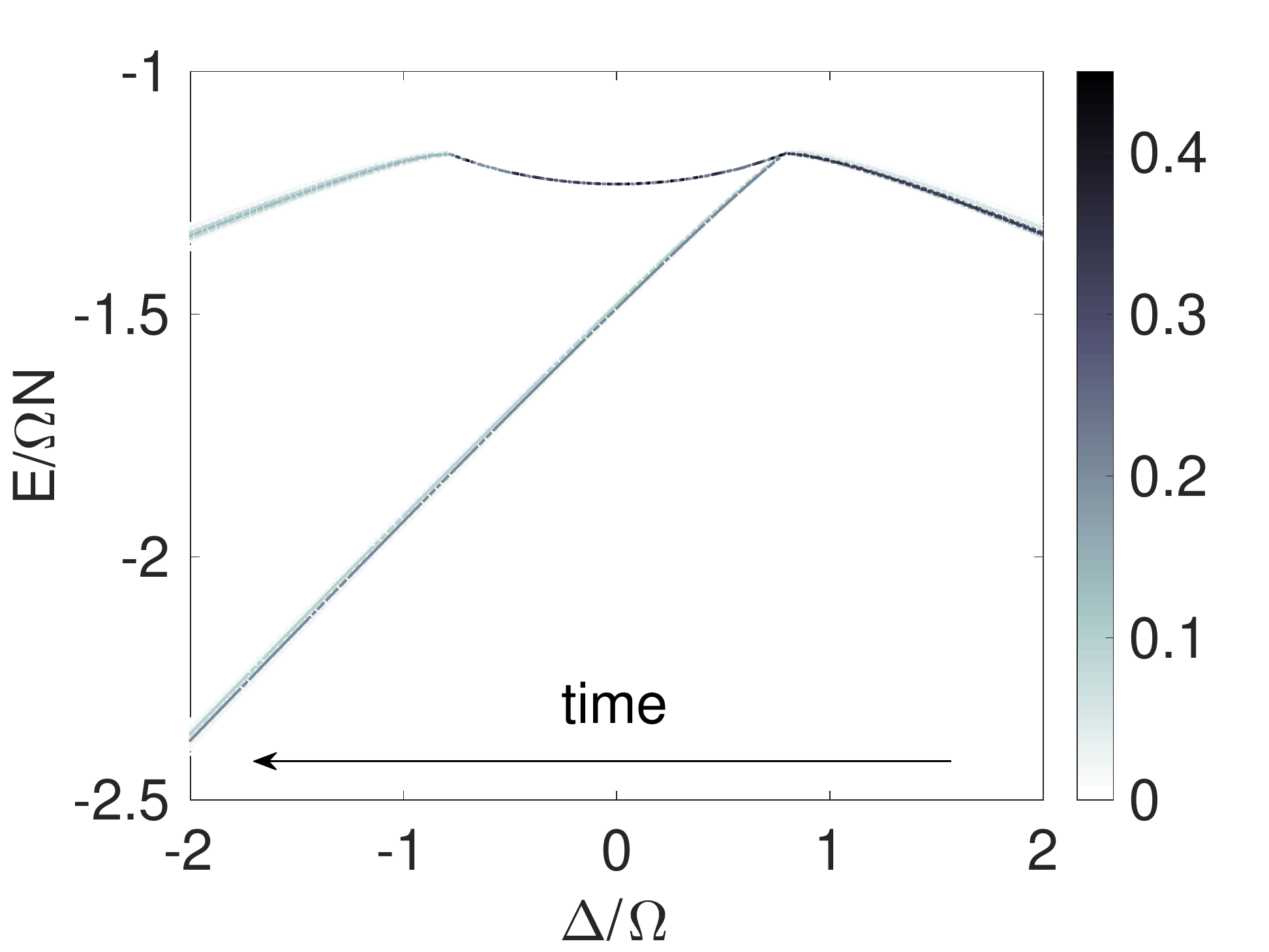}}
\subfloat[Classical]{\includegraphics[width=.32\textwidth]{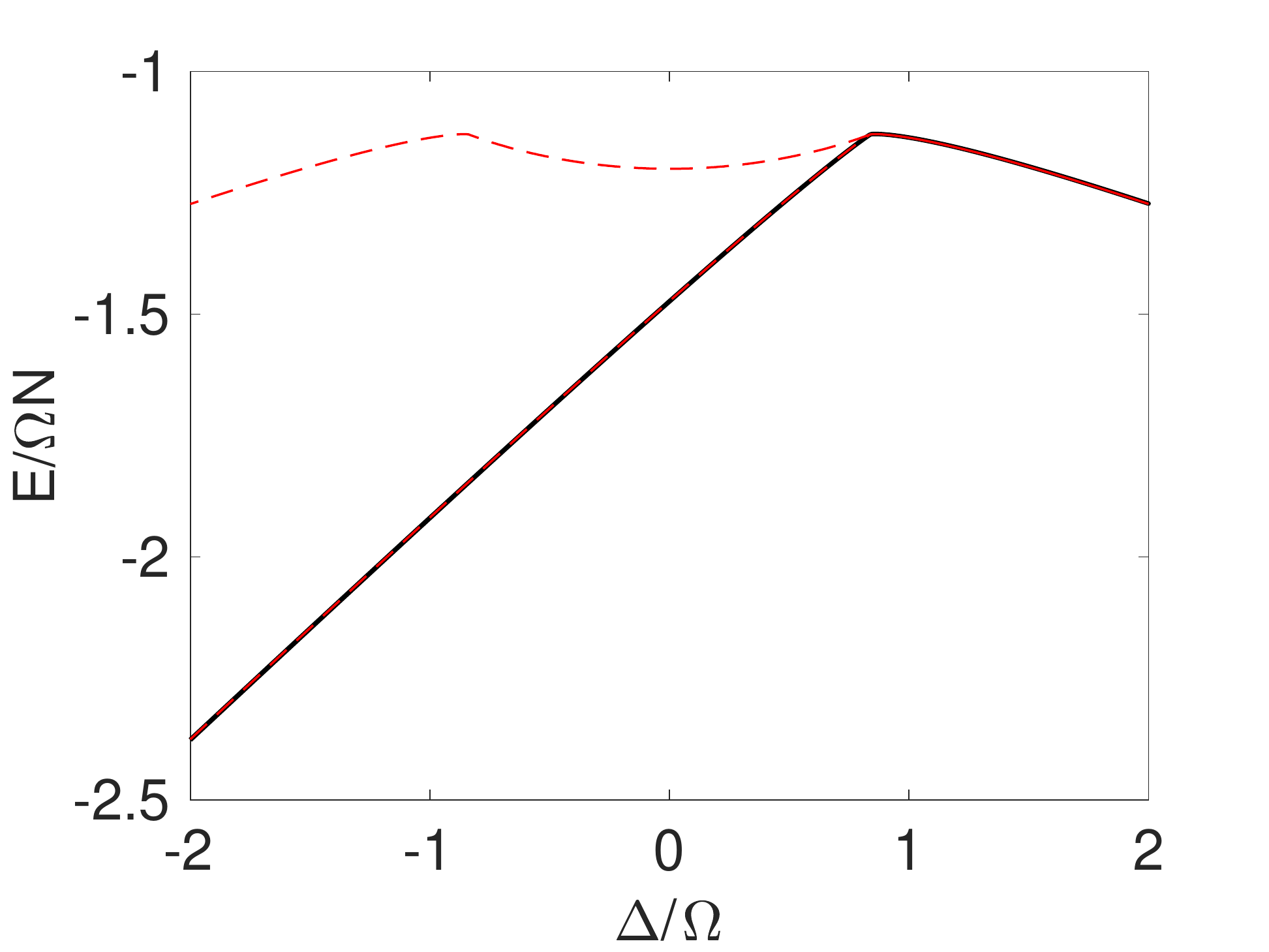}}
\caption{Classical regime: Panels (a) and (b) show the probability to find the system with $N=1000$ in the adiabatic eigenstate with energy $E$ during the forward and backward sweep with $T=5000\Omega^{-1}$. For comparison panel (c) shows two representative trajectories of the semiclassical ensemble with the same initial energy, one that returns to the initial state (solid black) and one that does not (dashed red). Note that in contrast to previous figures panel (c) here shows the forward and backward sweep in a single plot. We do not show the adiabatic quantum spectrum in this case because the levels are too dense to be seen. For the quantum simulation we have chosen the 37th initial eigenstate as an example because it has approximately the same energy per particle as the initial state of Fig.~\ref{fig:N=30} and Fig.~\ref{fig:N=100}.}
\label{fig:N=1000}
\end{figure*}

For even larger particle numbers the final probability distribution does indeed consist of two well separated peaks, much as in the {semi}classical case, and the probability for intermediate energies becomes vanishingly small (see Fig.~\ref{fig:N=100}).
Further increasing the particle number reduces the spread in energy of the two branches; see the results for very large particle number $N=1000$ in Fig.~\ref{fig:N=1000}. We do not show the adiabatic spectrum in Fig.~\ref{fig:N=1000}, as with $N=1000$ it is simply too dense to identify individual levels. Instead in Fig.~\ref{fig:N=1000}~(c) we show for comparison the energy of two trajectories of the semiclassical ensemble with the same initial energy and sweep rate. 
Qualitatively the result for $N=1000$ is very similar to the result for $N=100$, but the widths of the two final branches of the probability become narrower, and their energies are very close to the energies of the sets of trajectories of the semiclassical ensemble. 

\begin{figure}
\centering
\subfloat[$N=100$]{\includegraphics[width=.24\textwidth]{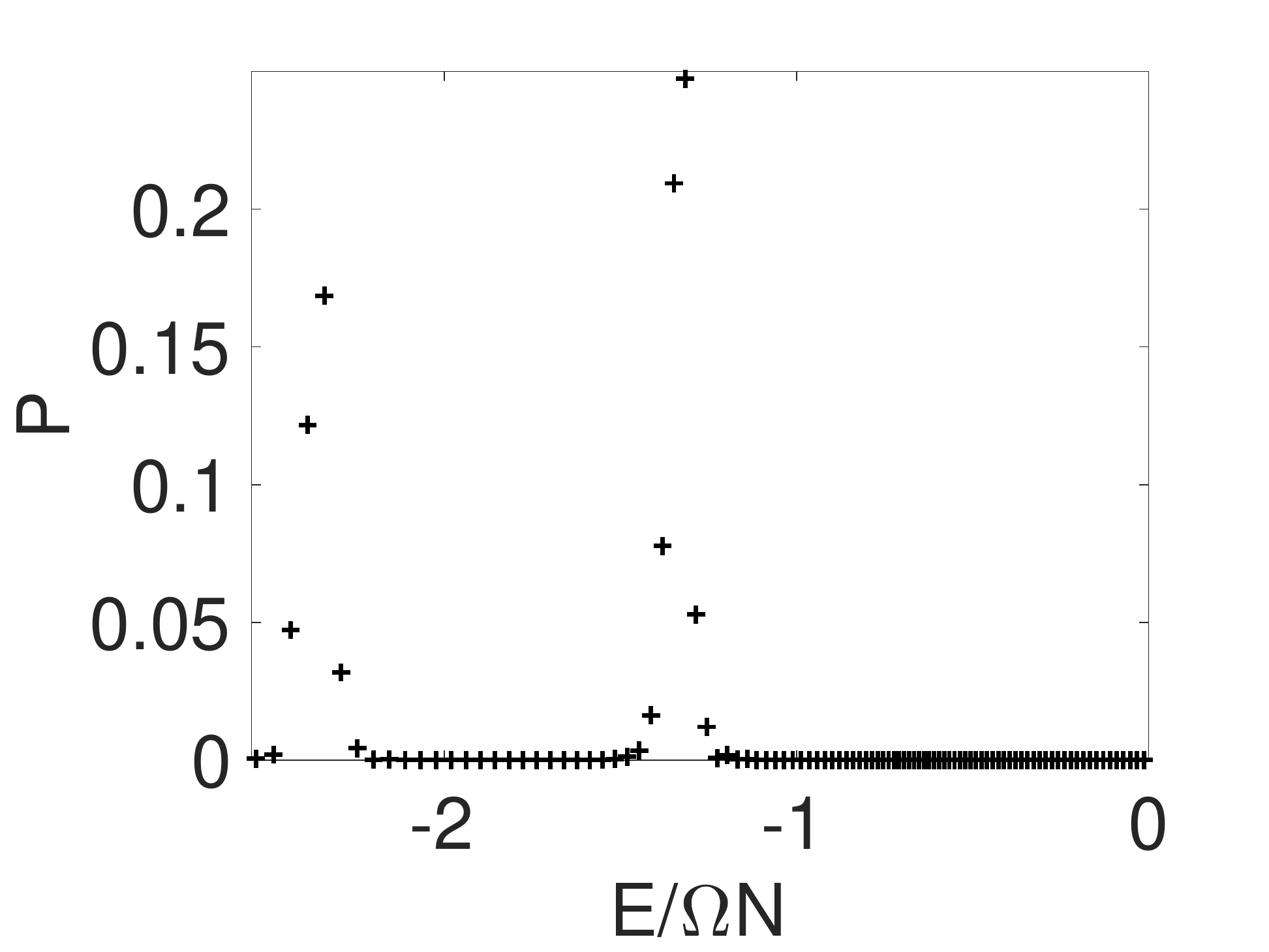}}
\subfloat[$N=1000$]{\includegraphics[width=.24\textwidth]{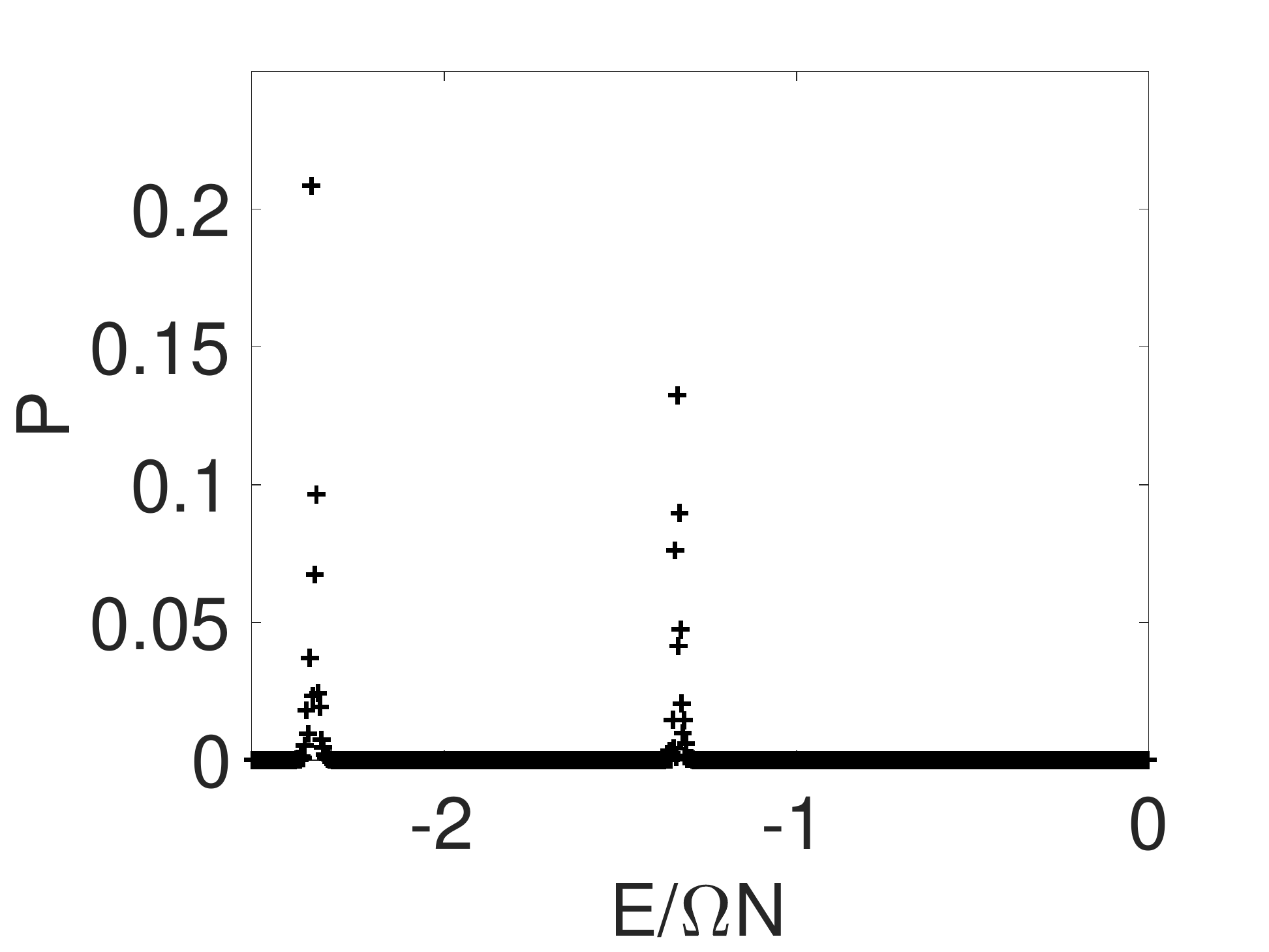}}
\caption{Final probability distribution at the end of the forward-and-back sweep for the same scenarios as in Fig.~\ref{fig:N=100} and Fig.~\ref{fig:N=1000}. There are two qualitatively different groups of final states: one with energy close to the initial state and one with much higher energy. Increasing the particle number reduces the widths of each of these two groups, so that as in the semiclassical case two sharp final energies are obtained for large $N$. The return probability $P_{\mathrm{ret}}$ is then defined as the sum of the probabilities in the left peak, which is centered around the initial energy.}
\label{fig:final_p}
\end{figure}

Fig.~\ref{fig:final_p} shows the final probability distribution of energies at the end of the sweep for $N=100$ and $N=1000$, \emph{i.e.} a slice along the vertical axis at $\Delta/\Omega=-2$ in Fig.~\ref{fig:N=100}~(c) and Fig.~\ref{fig:N=1000}~(b). 
As in the classical case there are now two qualitatively different outcomes of the experiment, corresponding to the inner disk and outer shell in \cite{dimer} or the inner and outer shell in Fig.~\ref{fig:phase_space}. In the quantum evolution this simply corresponds to two well separated peaks in the final probability distribution of energy eigenvalues. In the correspondence regime (iii) we can therefore unambiguously define the quantum return probability $P_{\mathrm{ret}}$ as the sum of the probabilities within the first peak. In regime (ii) such a definition would not be possible since the two peaks overlap, making a clear separation into qualitatively different fractions impossible. 

This unambiguous definition of $P_{\mathrm{ret}}$ in regime (iii) as the probability in the first final energy peak is also clearly measurable experimentally. Because the large final detuning completely dominates the tunneling term in the Hamiltonian, the final-time energy eigenstates are essentially eigenstates of particle number in the two sites. The two peaks in Fig.~\ref{fig:final_p} correspond to particle number distributions which are easily distinguished, even with very coarse particle-counting resolution. In this regime of clearly separated final energy peaks, therefore, it should be relatively straightforward to recognize, in each run of an experiment, which of the two possible final energy ranges has been reached in that run. After many runs the return probability can therefore be measured empirically in a straightforward manner.

\subsection{Quantitative correspondence?}
So far we have confirmed numerically that for large particle numbers, and for a sufficiently but not excessively slow sweep rate, \emph{qualitatively} similar results to the semiclassical results are obtained, inasmuch as an initial energy eigenstate evolves through the forward-and-back sweep into two narrow ranges of energies, each range well separated from the other. Probabilities for any final energies between these two distinct narrow ranges become extremely small. The probability to return to the initial energy range is theoretically well defined and experimentally measurable.

We have not yet fully confirmed the emergence of semiclassical probabilistic hysteresis from quantum mechanics, however, because the classical return probability is determined quantitatively by Kruskal's theorem, but we have not yet evaluated the quantum return probability quantitatively. Quantitative comparison is necessary even just to estimate the correct sweep time scale at which the semiclassical quasi-static result should be expected, since in the quantum quasi-static limit of infinite slowness the return probability will always be one. We have now seen that for some quantum cases, with large but finite $T$, the return probability is at least something less than one. In what range of sweep time scales $T$, if any, will the classical and quantum probabilities actually agree?

At least for particle numbers up to around $N=1000$ we can answer this question numerically for any particular case, by numerically solving the time-dependent many-body Schr\"odinger equation for the two-site Bose-Hubbard system. We can also make steps toward understanding the numerical results analytically, by considering the evolution in terms of quantum adiabatic theory. Because our sweep is slow, the quantum system certainly follows an adiabatic eigenstate until an avoided crossing is encountered. At every avoided crossing the probability for a diabatic transition can be calculated by the Landau-Zener formula. The whole sweep can thus be considered as a long series of Landau-Zener ``mini-sweeps'' through a succession of avoided crossings. 

\section{Quantum-Classical correspondence via Landau-Zener transitions}
\label{sec:LZ}
The celebrated Landau-Zener formula states that the probability for a diabatic transition over a parametric sweep through an avoided crossing of two energy levels $E_{1,2}$ is given by
\begin{equation}
P_{\mathrm{diab}}=e^{-\frac{2 \pi v^2}{|\alpha|}} \label{eq:LZ}
\end{equation}
where $v$ is the off-diagonal matrix element coupling the two levels and $\alpha=(E_2(t)-E_1(t))/t$ is the slope of the separation of the diabatic energy levels, provided that this separation changes linearly with time.  Although the Landau-Zener formula is derived for a two-level system, it is well known that it also applies to multi-state problems as long as the avoided crossings are well enough separated to justify a two-level approximation locally. This approximation is known as the \textit{independent crossing approximation} (ICA) \cite{Sinitsyn1, Sinitsyn2, Sinitsyn3, Sinitsyn4,Korsch}. 

The ICA is justified as long as adiabatic evolution only breaks down at any given time within orthogonal two-state subspaces of the total Hilbert space, so that there is never any need to compute non-adiabatic evolution within a subspace of dimension three or more, and the full evolution can be given as a tensor product of adiabatic and Landau-Zener evolutions. For example if one level has a narrowly avoided crossing with another level, and then later has another avoided crossing with another level, the two successive crossings may be treated independently as long as the ranges of $\Delta$ within which each crossing gives non-adiabatic evolution do not overlap. The ranges of $\Delta$ within which adiabaticity fails are defined by $E_{n+1}(\Delta)-E_n(\Delta)\lesssim \dot \Delta/\Delta$, and hence they can be made arbitrarily narrow by reducing the sweep rate. The ICA is therefore bound to be valid for slow enough sweeps. In the following sections we will discuss the ICA for our system.

\subsection{Independent crossing approximation (ICA)}
The ICA has been given several somewhat different implementations in the literature, which we discuss in the Appendix. We use here the variant in which the probability of a diabatic transition from the adiabatic level $i$ to $j$ at an avoided crossing is given as 
\begin{equation}
P_{ij}=e^{- \frac{\pi \delta_{ij}^2}{2 \dot \Delta \alpha_{ij}}},
\label{eq:imp_mod_ICA}
\end{equation}
where $\delta_{ij}$ is the size of the energy gap at the avoided crossing, $\dot \Delta=(\Delta_0-\Delta_I)/T$ is the sweep rate, and $\alpha_{ij}$ is the difference of the asymptotic slopes of the adiabatic levels $i$ and $j$. Further details concerning the method, and a comparison of the results obtained by successive application of (\ref{eq:imp_mod_ICA}) to a numerical solution of the Schr\"odinger equation, can be found in the Appendix. 
The ICA becomes a more accurate approximation for lower particle numbers (since the lower density of levels means fewer avoided crossings within any $\Delta$ range) and slower sweep rates (since the $\Delta$ ranges in which diabatic transitions occur become narrower). Very low $N$ or large $T$ are by no means needed, however; we show in the Appendix that the ICA remains good even for $N=1000$ and $\dot \Delta=8\cdot 10^{-4}$, which are the highest particle number and sweep rate that we have considered.

Note that (\ref{eq:imp_mod_ICA}) gives only the transition probabilities but no phase information, even though the Landau-Zener problem can be solved exactly for the complex transition amplitude including a phase. The phases of individual Landau-Zener transitions can only affect the final probability distribution, however, if there are interference effects between multiple transitions, as levels cross and re-cross. We will see that such interference affects can indeed occur; the reason to ignore the Landau-Zener phase is not that it never matters. The reason to ignore the Landau-Zener phase in the ICA is that if interference between multiple crossings is important then the crossings are not really independent, even though they are all nicely separate, and so since one has to keep track of a complicated array of many quantum phases anyway, one might as well just solve the whole problem numerically and forget the ICA.

By comparing such full numerical solutions with approximations based on the probabilities (\ref{eq:imp_mod_ICA}), however, we will see below that there are indeed conditions under which one may apply the ICA incoherently, summing probabilities given by (\ref{eq:imp_mod_ICA}) without considering phases. We will also find that it is precisely under these conditions that the correspondence of the quantum evolution with classical probabilistic hysteresis emerges. Before proceeding to these comparisons, we will first analyze the behavior of the Landau-Zener probabilities for individual crossings as given by (\ref{eq:imp_mod_ICA}).
 
\subsection{Crossover from diabatic to adiabatic}
As we discussed in Section V above, the energy gaps $\delta_{ij}$ at most of our avoided crossings become extremely small in number regimes (ii) and (iii), which only require $N\gtrsim 20$. Consequently, for realistic sweep rates and not too small particle numbers, many of the probabilities given by (\ref{eq:imp_mod_ICA}) tend to be close to one. In particular, if the system starts in a low-energy state then the probabilities for diabatic transitions through the first avoided crossings that are encountered, as $\Delta(t)$ is initially swept forward, are all close to one. 

As the sweep continues through its succession of diabatic avoided crossings, however, the system is steadily moving upwards in the $(E,\Delta)$ plane through the swallowtail region, towards the separatrix at the top of the swallowtail. Beyond the separatrix there are no more {extremely narrow} avoided crossings; all energy gaps are of order $\Omega$. As the system moves up through the swallowtail to approach the separatrix, therefore, the energy gaps $\delta_{ij}$ at the avoided crossings must eventually begin to be larger.

\begin{figure}
\includegraphics[width=0.45\textwidth]{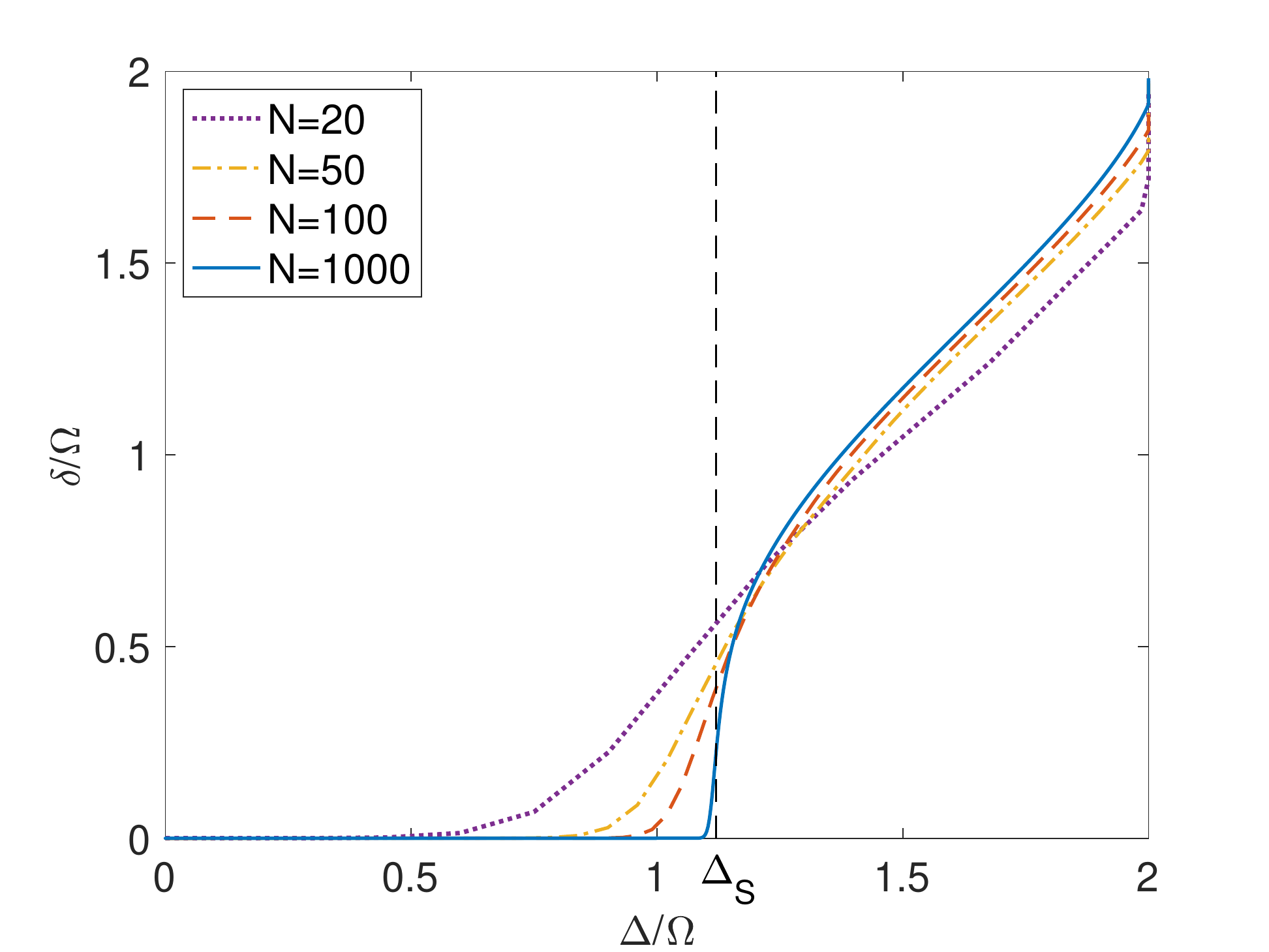}
\caption{Energy gaps $\delta$ at avoided crossings encountered during the sweep, for various particle numbers $N$, if the initial state is the ground state and the system is assumed to follow the diabatic path at every avoided crossing. For visual clarity the discrete sequences of encountered crossings, each with a particular $\delta$ at a particular $\Delta$, are represented by interpolation as continuous curves $\delta(\Delta)$. In all cases the gaps at the first avoided crossings are extremely small, so that the system indeed takes the diabatic path with extremely high probability. The gaps become larger around $\Delta=\Delta_S$, which is the value of $\Delta$ at which the classical separatrix shrinks to a point and vanishes (dashed line at $\Delta/\Omega\sim 1.1$). Over a small range of $\Delta$ around $\Delta_S$, $P_{\mathrm{diab}}$ drops to very low values, and thereafter the system essentially just follows the adiabatic path. For larger $N$ the transition from diabaticity to adiabaticity around $\Delta_S$ becomes more abrupt, but the the density of the level spectrum also increases with $N$, and for large $N$ there are many crossings within the range of $\Delta$ over which the avoided crossing gaps grow. There are thus many avoided crossings within this range for which the evolution is between diabatic and adiabatic, with significant Landau-Zener probability to take each branch. Through each such crossing the system's quantum state is split coherently into a superposition of the two branches.}
\label{fig:spacings}
\end{figure}

Fig.~\ref{fig:spacings} shows this pattern by showing the sequence of $\delta_{ij}$ that would be encountered in succession on the forward sweep, if the system began in the ground state and then went through every crossing diabatically. Each successive $\delta_{ij}$ would be a single point in the $(\Delta,\delta)$ plane, but we wish to compare cases with greatly differing $N$, for which the points would be differently spaced because for larger $N$ the avoided crossings are packed more closely together. Fig.~\ref{fig:spacings} therefore shows the $\delta_{ij}$ sequences for all $N$ as interpolated smooth curves $\delta(\Delta)$. The curves are also extrapolated to show large avoided crossing widths $\delta\sim\Omega$ at larger $\Delta>\Delta_{S}$ even though the ``crossings'' in this range are really just energy differences between levels that never cross. 

In spite of these subtleties in interpreting Fig.~\ref{fig:spacings}, its implications should be clear. The point $\Delta=\Delta_S$ where the quantum $\delta_{ij}$ suddenly rise for large $N$, when the initial state is the ground state, is precisely the point at which adiabaticity breaks down classically even in the quasi-static limit, allowing quasi-static irreversibility to occur \cite{dimer}. A similar pattern occurs for other initial states; the general phenomenon is that as the quantum system approaches the separatrix in the $(E,\Delta)$ plane, the avoided crossing gap $\delta$ begins suddenly climbing. For large $N$ the climb is almost vertically sharp and begins abruptly right at the separatrix, with $\delta$ remaining very small until just below the separatrix and then already being of order $\Omega$ just above it. (This correspondence between the classical separatrix and abrupt increase in quantum level splittings has previously been noticed in \cite{Aubry}.)

The classical result of perfect reversibly in cases where $\Delta$ reverses its sweep before the initial ensemble meets the separatrix is thus also clear quantum mechanically. For evolution that never brings the quantum system to the $(E,\Delta)$ separatrix, all the energy gaps encountered during the sweep are vanishingly small for even moderate $N$, so the system follows the diabatic path with essentially unit probability in both the forward and backward sweep,  and the return probability is essentially one. The system essentially remains at all times in a single adiabatic energy state. 

What of quantum-classical correspondence in cases where the $\Delta$ sweep continues through $\Delta_{S}$, though?

\subsection{Between diabatic and adiabatic}
Insofar as there even are any avoided crossings above the quantum separatrix, they are wide enough for the Landau-Zener evolution through them to be essentially perfectly adiabatic. Below the quantum separatrix, on the other hand, the avoided crossings are so narrow for any practical sweep rate, even for moderate $N\gtrsim50$, that their Landau-Zener evolution is essentially perfectly diabatic.  Although the crossover is quite sharp for large $N$, from nearly perfect quantum diabaticity below the separatrix to nearly perfect adiabaticity above it, the transition from diabaticity to adiabaticity is not fully accomplished between one avoided crossing and the next, not even in dynamical regime (iii).

For large $N$ the avoided crossings are densely packed along $\Delta$, and as the system approaches the separatrix there are many avoided crossings with $P_{\mathrm{diab}}$ neither close to zero nor to one. Through each of these intermediate avoided crossings the system's state bifurcates significantly into a superposition of two different energy eigenstates. Consequently, the quantum state after the sequence of many of these transitions around $\Delta_{S}$ is a coherent superposition of many adiabatic eigenstates. If the $\Delta$ sweep continues further, subsequent avoided crossings under classically adiabatic conditions $\Omega T\gg1$ are all essentially perfectly adiabatic, so no further bifurcations occur on the forward sweep. 

On the backward sweep through the same set of avoided crossings, all these quantum amplitudes interfere at each crossing, providing even more non-trivial bifurcation. If a single energy eigenstate is coherently split by a Landau-Zener transition of intermediate probability, then passing back through the same Landau-Zener transition can potentially merge the two branches of the superposition back into the single initial eigenstate---but only if the two branches begin the reverse transition with exactly the right relative phase. If we were dealing here with only a single avoided crossing of two states, then we could achieve this perfect reversal of a bifurcation just by carefully timing our sweep so that the accumulated adiabatic phases of the two branches had the right difference. With amplitude spread over many energy eigenstates, however, and then returning through a sequence of many non-trivial avoided crossings, there is no way to ensure that the initial single energy eigenstate is recovered with unit probability.

The quantum forward-and-back sweep process for $N \to \infty$ thus exhibits probabilistic hysteresis for $\Delta_0>\Delta_S$, just as  the classical process also does, unless the quantum sweep is so impossibly slow that even the exponentially narrow avoided crossings are adiabatic. Such slowness is only remotely feasible in dynamical regime (i) ($N\lesssim 20$ for $u=-3$); already in regime (ii) ($20\lesssim N\lesssim 50$ for $u=-3$) there is quantum probabilistic hysteresis. In regime (ii) the Landau-Zener $P_{\mathrm{diab}}$ may range between 0 and 1 for avoided crossings within the full area of the $(E,\Delta)$ swallowtail, so that bifurcations leading to probabilistic hysteresis are not only associated with the separatrix itself, as they are in the classical problem. This regime is thus complex and we will analyze it no further in this paper.

In regime (iii) ($N\gtrsim 50$ for $u=-3$) the avoided crossings are all extremely narrow except very close to the separatrix, and so bifurcations into many energy eigenstates only occur within that narrow region. Because the separatrix region has high density of states, these many eigenstates are all within a narrow range in energy, even though they are many. For slow sweeps at the large $N$ of regime (iii), therefore, the quantum system becomes nonadiabatically spread over many states within a narrow range of energies right around the classical separatrix. Under subsequent adiabatic evolution this narrow range separates into two narrow ranges that move quite far apart from each other in energy, much as the classical ensemble was likewise divided in \cite{dimer}.

This is still only a qualitative quantum-classical correspondence, however, inasmuch as both cases show some degree of probabilistic hysteresis. What circumstances can further ensure that the quantum and classical probabilities are actually the same?

\subsection{Correspondence failure}
\begin{figure}
\includegraphics[width=0.45\textwidth, trim=17mm 0mm 0mm 0mm, clip]{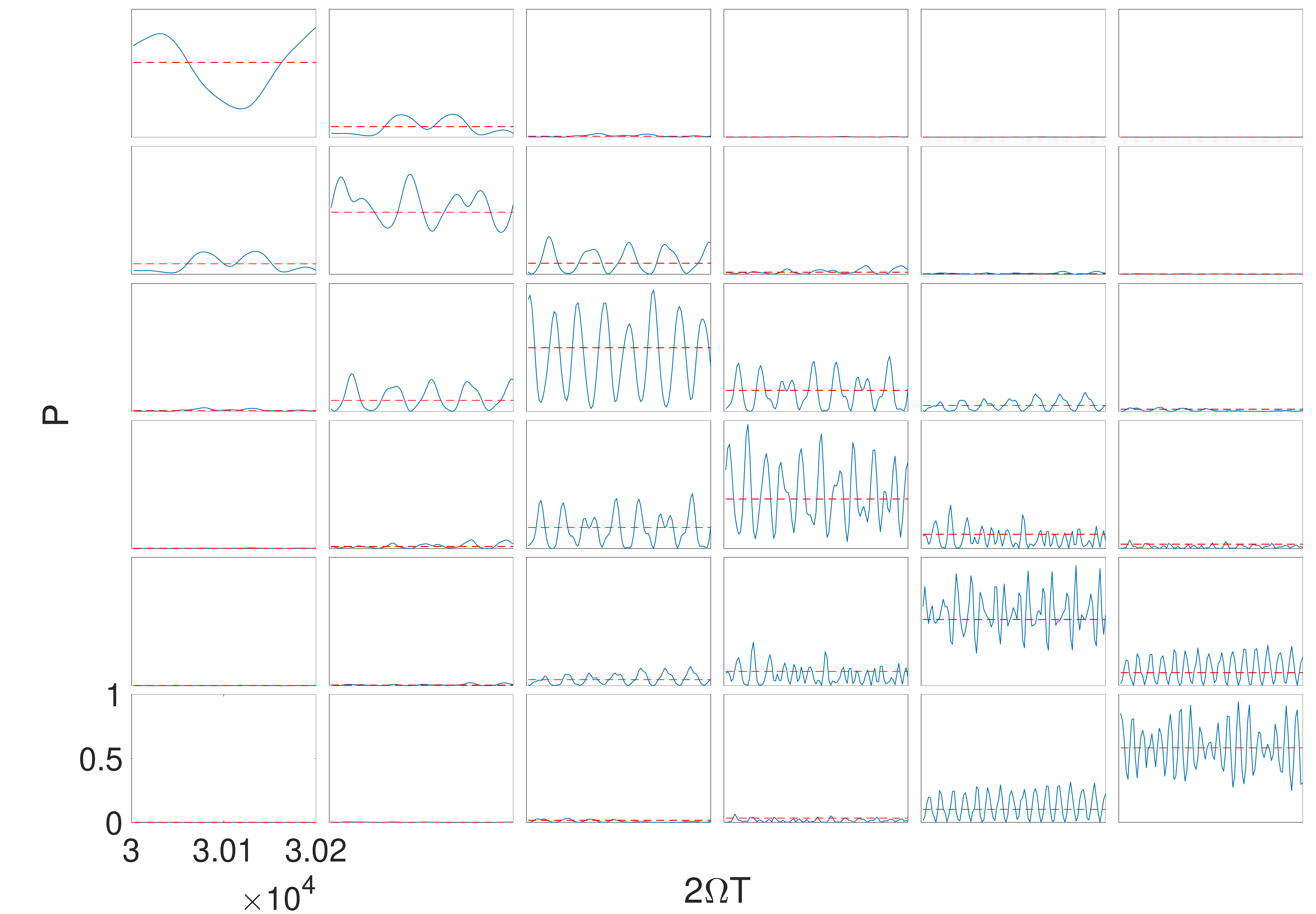}
\caption{Each panel shows the probability (solid blue) to end up in the eigenstate of the initial Hamiltonian that is given by the panel's column, provided that the initial state was the state given by the panel's row (i.e. the panel in the $i$-th column and $j$-th row shows the probability to be in the $i$-th state at the end of the sweep, starting initially in the $j$-th state). The dashed red line shows the corresponding probability in the incoherent Landau-Zener approximation which is basically constant in the displayed $T$ range. The incoherent Landau-Zener approximation gives the correct short-time average of the probability. This is also true for the panels far away from the diagonal, where the probabilities become too small to be seen in this figure. The system parameters are $N=50$, $u=-3$, $\Delta_I/\Omega=-2$ and $\Delta_0/\Omega=1$.}
\label{fig:incoherent}
\end{figure}

First of all we can confirm that the quantum and classical probabilities do \textit{not} automatically coincide for large $N$. To confirm this we consider several different eigenstates of the initial Hamiltonian $\hat H(t=-T)$ as initial states, and for each of them we numerically evolve through a forward-and-back sweep at a slow rate $1/T\ll\Omega$ to compute the probabilities of ending up in various final energy eigenstates at $t=+T$. What we can then see in Fig.~\ref{fig:incoherent} is that all these quantum probabilities depend sensitively on the precise sweep time scale $T$: variations in $T$ of less than one percent can change the quantum probabilities by factors of order unity, even when $\Omega T$ is very large. The semiclassical probabilities, in stark contrast, are independent of $T$ as long as $T\gg\Omega^{-1}$. 

The quantum variations in probability are in fact oscillatory as functions of $T$; this indicates that they are due to interference effects through the avoided crossing network, which are sensitive to time-dependent phase differences between bifurcated branches of the quantum state in the instantaneous energy basis. It is possible that the amplitude of these oscillations will eventually become small at some very large $N$, but our numerical investigations at the largest computationally feasible $N$ of around 1000 indicate that the approach to quantum-classical correspondence with increasing $N$ is very slow (possibly logarithmic). 

We emphasize that this sensitive dependence on the precise sweep rate is only revealed because we have included the backward sweep in our protocol. After only the forward sweep, when the amplitude has only been spread over a narrow range of energies, the differences the system state for different sweep rates are very subtle, in the sense that the probability distribution among nearby states is only slightly changed. In particular what is called ``Landau Zener tunnelling probability'' in \cite{Trimborn3, Liu, Wu, Wu2}, namely the ratio of the populations of the two sites at the end of the forward sweep, does \textit{not} sensitively depend on the precise sweep rate.

\subsection{The incoherent Landau-Zener approximation}
Also shown in Fig.~\ref{fig:incoherent}, however, are dashed lines indicating the probabilities for each final state, given each initial state, that we obtain if we apply the ICA at every avoided crossing during the forward-and-back sweep, but taking only the probabilities $P_{\mathrm {{diab}}}$, neglecting phases and all interference effects. We call this calculation the \textit{incoherent Landau-Zener approximation}. The dashed curves that it produces in Fig.~\ref{fig:incoherent} are all essentially horizontal lines, with negligible dependence on the precise sweep time scale $T$. This is because ignoring phases and interference is equivalent to time-averaging all the phases, and then since the phases evolve quite quickly while the sweep of $\Delta(t)$ is slow, time-averaging is essentially equivalent to averaging over the sweep time scale $T$, since slightly faster or slower sweeps lead to greatly differing relative phases accumulating over the times between successive avoided crossings, {whereas $P_{\mathrm{diab}}$ changes on a much longer time scale}. The incoherent ICA thus effectively averages away the oscillations of probabilities with sweep rate $T$, yielding average probabilities that are independent of $T$.

The incoherent Landau-Zener approximation will thus be an accurate approximation for experiments in which return probabilities are measured by repeated runs, with imperfect control over variations in sweep time $T$ among runs. In this sense the averaged probability given by the incoherent Landau-Zener approximation emerges naturally as the quantity that can most straightforwardly be measured.

A second sense in which the incoherent ICA is actually realistic appears when we note in Fig.~\ref{fig:incoherent} how the probabilities for any single sweep rate $T$ vary with starting eigenstate $i$. Especially for large particle numbers $N$ it is experimentally unrealistic to prepare a single eigenstate initially; more realistic is a microcanonical or canonical ensemble (as in \cite{dimer}) with a finite energy width. Experimental measurements of return probability will therefore effectively sum over initial eigenstates $i$, even for a single run with a fixed $T$. When many adiabatic eigenstates participate in the evolution, the oscillations of probability shown in Fig.~\ref{fig:incoherent} are smeared out \cite{Wilkinson}, effectively reproducing the incoherent ICA result, as we demonstrate below. Since the dimension of the Hilbert space is $d=N+1$, the average number of eigenstates in any normalized energy range $\Delta E/\Omega N$ is proportional to $N$, so that even for a very small energy width a large number of levels is initially occupied, for large enough $N$. 

\begin{figure}
\centering
\subfloat[single state]{\includegraphics[width=.24\textwidth]{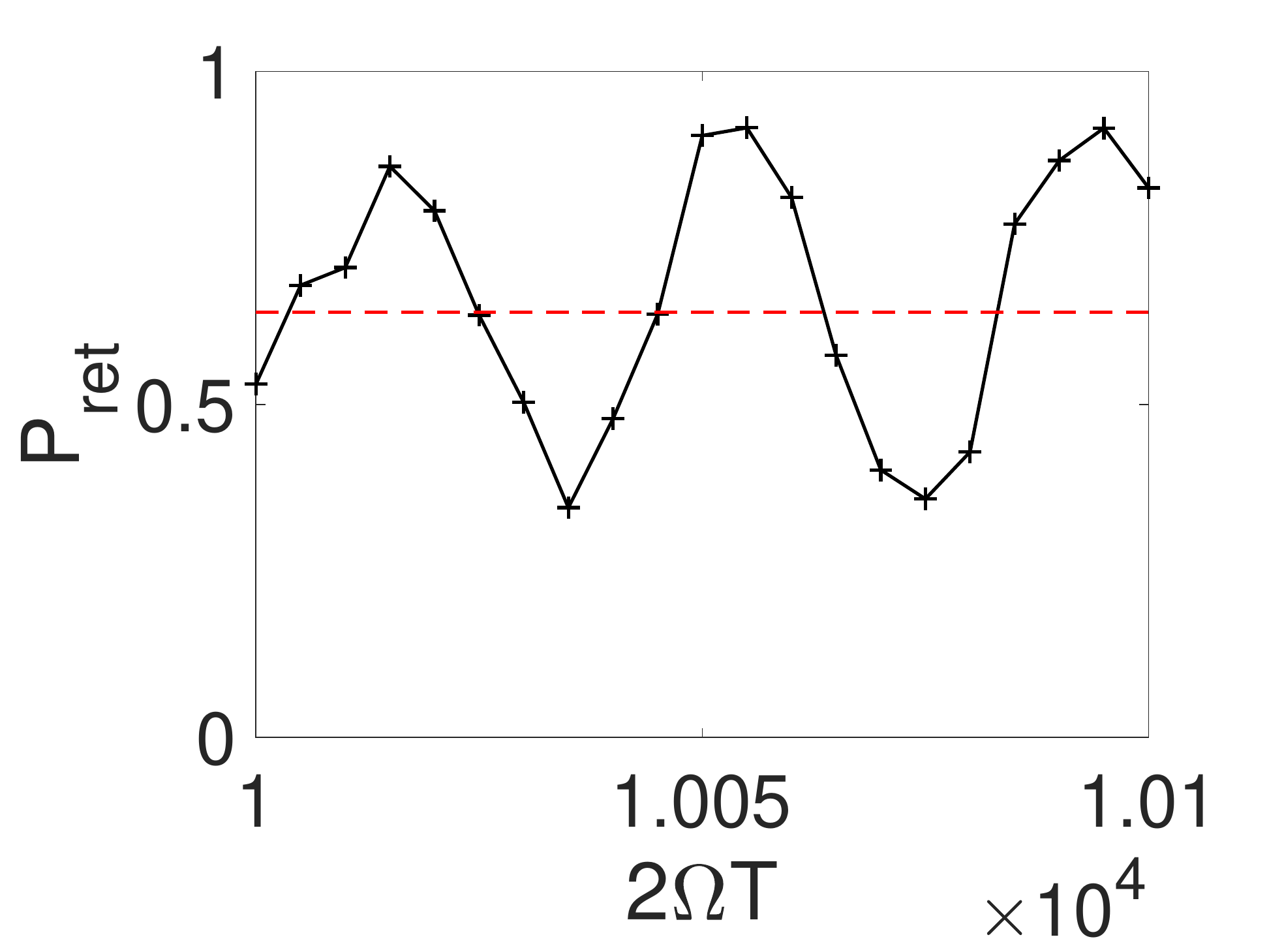}}
\subfloat[ensemble]{\includegraphics[width=.24\textwidth]{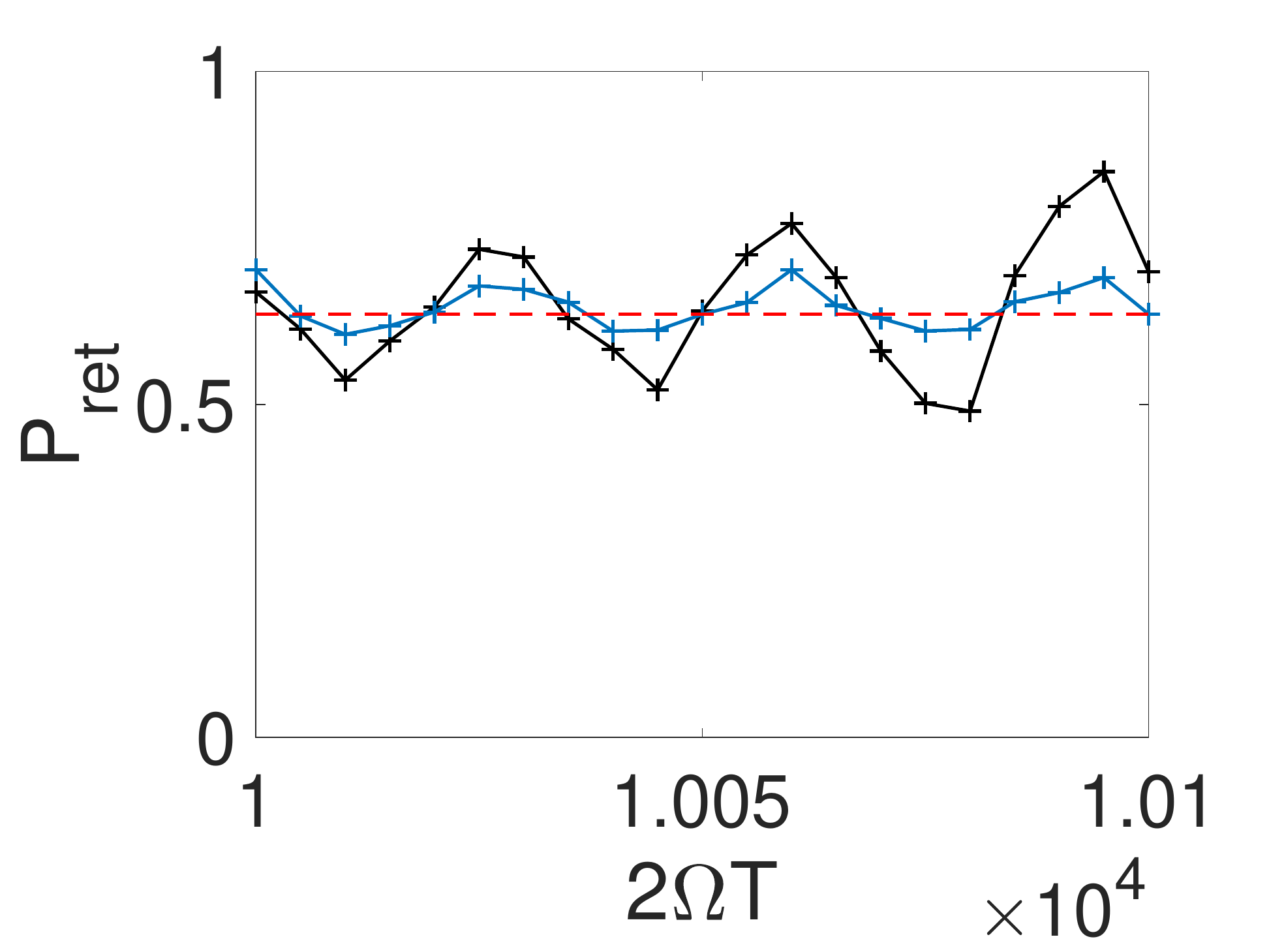}}
\caption{Comparison of the return probability (solid) with the incoherent Landau-Zener approximation (dashed red) for different sweep times and different initial states (a: 37th eigenstate; b: ensemble containing four (black) or twenty (blue) eigenstates with the same mean energy as in (a)). The oscillations around the incoherent Landau-Zener result are strongly suppressed for the finite width ensembles in the right panel. Since even just a few states are enough to suppress the oscillations, the energy width can still be very small. Note that the incoherent Ladau-Zener result itself changes by less than $10^{-5}$ in the displayed $T$ range.  The system parameters are $N=1000$, $u=-3$, $\Delta_0/\Omega=-\Delta_I/\Omega=2$.}
\label{fig:p_N=1000}
\end{figure}

Fig.~\ref{fig:p_N=1000} shows the return probability for $N=1000$ and a single initial state (left) and two initially microcanonical ensembles with different energy width (right) over time. The dashed line indicates the incoherent Landau-Zener result.
We find that the oscillations in the return probability are suppressed in the case of a finite width ensemble so that even when the sweep time is held exactly constant between different runs the incoherent Landau-Zener approximation gives the correct return probability. For even larger particle numbers or initial energy range we expect the oscillations to vanish completely, as in the semiclassical case \cite{dimer}.

\subsection{Recovering quantum-classical correspondence}
\begin{figure}
\includegraphics[width=0.45\textwidth]{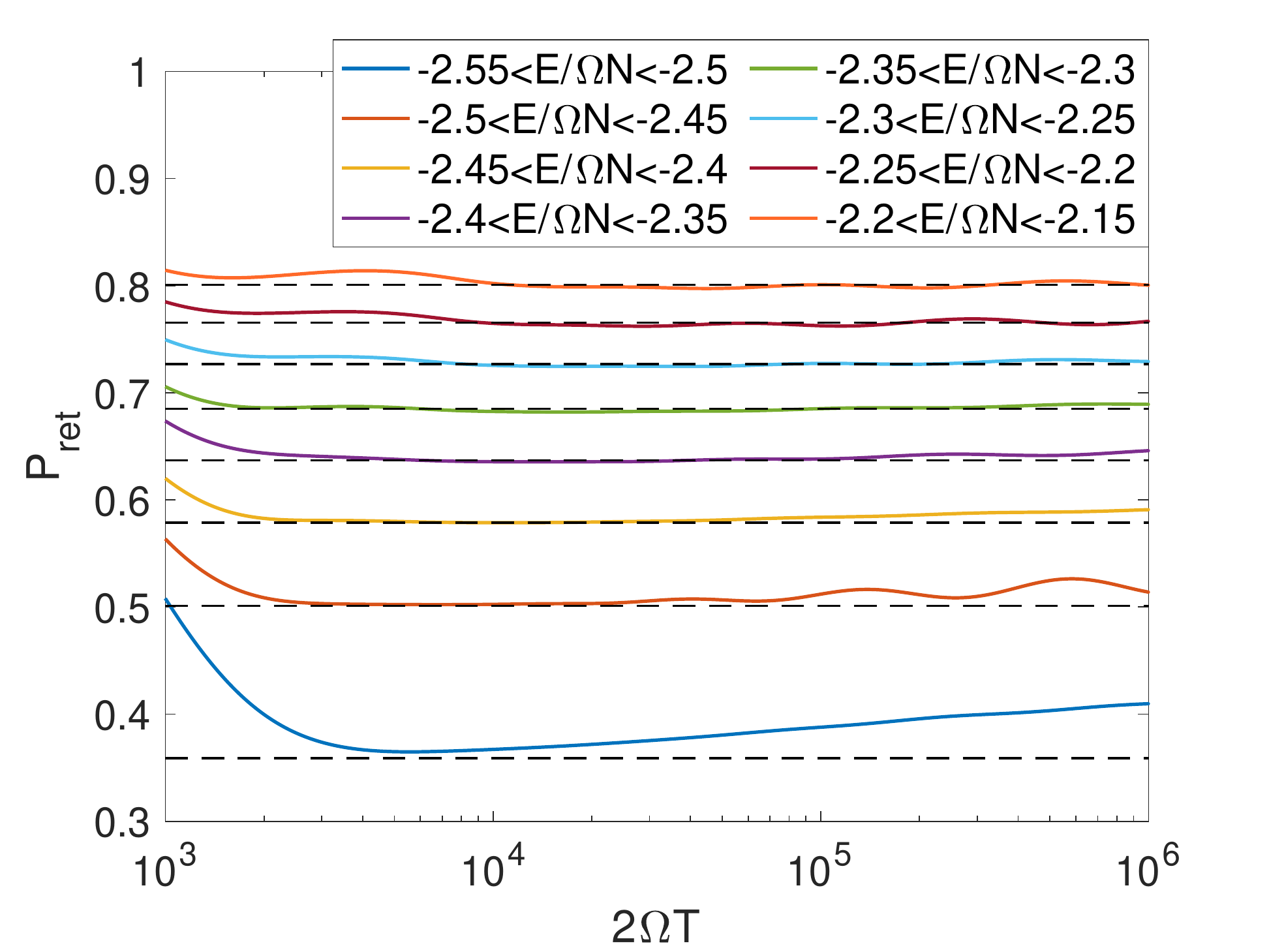}
\caption{Comparison of the return probability in the incoherent Landau-Zener approximation for microcanonical initial ensembles (colored lines) and the corresponding semiclassical values in the quasi-static limit (dashed). Note that the logarithmic horizontal axis covers a wide range of sweep times. Since it has been demonstrated in Fig.~\ref{fig:p_N=1000} that the incoherent Landau-Zener approximation does indeed closely approximate the quantum return probability, we conclude that the quantum return probabilities for initial quantum ensembles agree well with the classical values over a broad range of sweep rates. The system parameters are $N=1000$, $u=-3$, $\Delta_0/\Omega=-\Delta_I/\Omega=2$.}
\label{fig:qc_comparison}
\end{figure}

After having thus established the incoherent Landau-Zener approximation as accurate for realistic experiments with imperfectly reproducible $T$ {or, more importantly,} finite initial energy width, we can finally compare the quantum results for the return probability calculated with the incoherent Landau-Zener approximation with the semiclassical results. Fig.~\ref{fig:qc_comparison} shows the return probability for $N=1000$ plotted against the total sweep time $2T$ for initially microcanonical ensembles of states with different energy. The dashed lines indicate the corresponding semiclassical adiabatic values.
We find that there exists a broad range of sweep rates for which the quantum return probability is approximately constant and agrees with the semiclassical adiabatic value found in \cite{dimer}. The existence of this correspondence regime of probabilistic hysteresis is on the one hand surprising because it is not the quantum adiabatic limit (where the return probability would be always one, as discussed above), and yet the return probability is nearly independent of the sweep rate. On the other hand the existence of this correspondence regime is expected, since ultimately the semiclassical results must emerge from the full quantum description. For very slow sweep rates and low initial energy, a significant deviation from the semiclassical results can still be observed; the semiclassical result seems to underestimate the quantum return probability in these cases. The range of sweep rates over which quantum-classical correspondence is good for probabilistic hysteresis is quite broad, however, even at these lowest energies. That is our paper's main result.

We emphasize both that the incoherent Landau-Zener result corresponds to the semiclassical result and that a finite initial energy width is needed to justify {this} incoherent treatment. In principal we can obtain very similar results as in Fig.~\ref{fig:qc_comparison} also for smaller particle numbers ($N\sim 100$), but then the initial energy width has to be very wide to support enough quantum states (e.g. $\Delta E/\Omega N\sim 0.25$ for $N=100$ so that we would have only two lines in Fig.~\ref{fig:qc_comparison}). If the energy width is too small, the return probability oscillates around the semiclassical value---a purely quantum effect. We therefore conclude that even for very large $N$ strong quantum effects can be observed if the initial energy width is small enough (see Fig.~\ref{fig:p_N=1000}).

\section{Discussion}
\subsection{Summary and conclusion}
In conclusion we have demonstrated how irreversibility in the form of probabilistic hysteresis emerges in a small isolated quantum system and how the semiclassical limit is attained. Instead of the separatrix crossing mechanism discussed in \cite{dimer} for the classical system, the origin of irreversibility in the quantum system was found to be in a series of Landau-Zener crossings, leading at large $N$ to two qualitatively different outcomes of the sweep experiment. Although in the true quantum adiabatic limit the evolution is always fully reversible,  we showed that already for modest particle numbers the sweep rate has to be unrealistically slow to reach this limit. For large particle numbers we found that the return probability as a function of the sweep rate is almost constant over nearly three orders of magnitude, and within this broad range of slow sweep rates it agrees closely with the semiclassical adiabatic prediction. While it is therefore technically true that the adiabatic and semiclassical limits do not commute, we have shown that there still exists a broad quantum regime in good correspondence with the semiclassical adiabatic limit. 

To obtain these results it is important to have a finite energy width initially, so that several quantum states are supported and quantum interference effects are averaged out. In this case the return probability can be calculated accurately by the incoherent Landau-Zener approximation. We therefore conclude that the semiclassical adiabatic limit corresponds to the incoherent sum of all Landau-Zener probabilities, with quantum interference effects neglected. If only a single state is initially occupied, however, the interference effects lead to significant oscillation of the return probability around the semiclassical value predicted by the incoherent Landau-Zener approximation, even for very large particle numbers. Therefore strong quantum effects can be observed at large particle numbers provided that the initial energy width is small enough. The semiclassical value of the return probability is obtained only if, in addition to being in the limit of large particle numbers, a finite initial energy width also applies. 

\subsection{Outlook}
The results in this paper are all based on the Landau-Zener description of the sweep process. There is, however, a complementary view point using the phase space formulation of quantum mechanics in terms of quasi-probability distributions such as the Wigner function or Husimi function. Since the semiclassical description is based on considerations in phase space we expect that these quantum phase space methods can give further insights into the relationship between the quantum and classical mechanics of probabilistic hysteresis as the microscopic limit of irreversibility. Details will be published elsewhere.

The results reported here have all been purely numerical, and in particular the main result of correspondence between the incoherent Landau-Zener approximation and the semiclassical probabilities derived from Kruskal's theorem has been obtained purely as a numerical fact. The implication of this fact is that the Landau-Zener transition probabilities in the quantum problem are somehow related to the classical rates of phase space area growth that are involved in Kruskal's theorem. It should therefore be possible to demonstrate this relationship analytically. We are pursuing this goal.

The system studied here is completely integrable. It has been shown in \cite{trimer}, however, that chaos has strong effects on microscopic irreversibility and probabilistic hysteresis. Instead of introducing non-integrability by adding another degree of freedom to the dimer system, perturbations in the form of periodic kicks of either the coupling constant $\Omega$ or the interaction parameter $U$ can lead to dynamical chaos. In classical  phase space a chaotic strip first forms close to the separatrix of the unperturbed system. This way of introducing chaos should also work in the quantum dimer presented here, where we expect the level spacing statistics to change from a Poissonian to a Wigner-Dyson distribution around the maximum of the density of states, so that the effect of quantum chaos in probabilistic hysteresis can in future be studied by fairly straightforwardly modifying our integrable Hamiltonian.

\acknowledgements
The authors acknowledge support from State Research Center OPTIMAS and the Deutsche Forschungsgemeinschaft (DFG) through SFB/TR185 (OSCAR), Project No. 277625399.

\appendix*
\section{Variants of the ICA}
There are several variants of the ICA; here we discuss them. The standard form for multi-state Landau Zener assumes a Hamiltonian of the form $\hat H=\hat A +\hat B t$, where $\hat A$ and $\hat B$ can be represented by constant Hermitian matrices. Our Hamiltonian (\ref{eq:H}) has this form if we treat the forward and backward sweep separately. The \textit{diabatic} basis is the basis in which $\hat B$ is diagonal, which in our case is the Fock basis $\ket{n_1,N-n_1}$. In this basis the Hamiltonian~(\ref{eq:H}) assumes the form
\begin{equation}
\hat H=\begin{pmatrix}\varepsilon_0+\beta_0 t& v_0 &0&\dots&0\\
v_0&\varepsilon_1+\beta_1 t& v_1 &\dots&0\\
\vdots&\vdots&\ddots&\vdots&\vdots\\
0&0&\dots&\varepsilon_{N-1}+\beta_{N-1} t&v_{N-1}\\
0&0&\dots&v_{N-1}&\varepsilon_{N}+\beta_{N} t\end{pmatrix} 
\end{equation}  
with
\begin{equation}
\begin{split}
\varepsilon_i&=U\left(\frac{N^2}{2}+i^2-Ni\right)\\
\beta_i&=\frac{\dot \Delta(t)}{2}(2i-N)\\
v_i&=\frac{\Omega}{2} \sqrt{(i+1)(N-i)}.
\end{split}
\end{equation}
The diabatic levels are then given by the diagonal elements of $\hat H$ (i.e. the eigenvalues of the uncoupled system $v_i=0$), and their separation depends linearly on $t$. The goal of the ICA is to treat each avoided level crossing as if it were an avoided crossing in a two-level system, and apply (\ref{eq:LZ}). To do so the corresponding parameters $v$ and $\alpha$ have to be found for every avoided crossing.

In the easiest form of the ICA, often used for problems where the ICA result can be calculated analytically, the slope $\alpha$ which is relevant for the transition between the levels $i$ and $j$ is assumed to be $\beta_j-\beta_i$, and the parameter $v$ is given by the matrix element $H_{ij}$ \cite{Sinitsyn1, Sinitsyn2, Sinitsyn3, Sinitsyn4}. This procedure is motivated by the two-level system, where $v$ is exactly half of the size of the energy gap $\delta$ at the avoided crossing. In the multi-state problem, however, $\delta/2$ and $v$ are in general different; in particular there can be indirect coupling of two diabatic levels even if the corresponding matrix element vanishes \cite{Korsch}. In our system there is only one non-zero off-diagonal matrix element for each diabatic level $i$ coupling it to the level $i+1$, but clearly there can be level transitions at every avoided crossing. These transitions can be captured by using $\delta/2$ instead of $v$ in the Landau-Zener formula. this better approximation was called \textit{modified ICA} in \cite{Korsch}.

\begin{figure}
\centering
\includegraphics[width=0.45\textwidth]{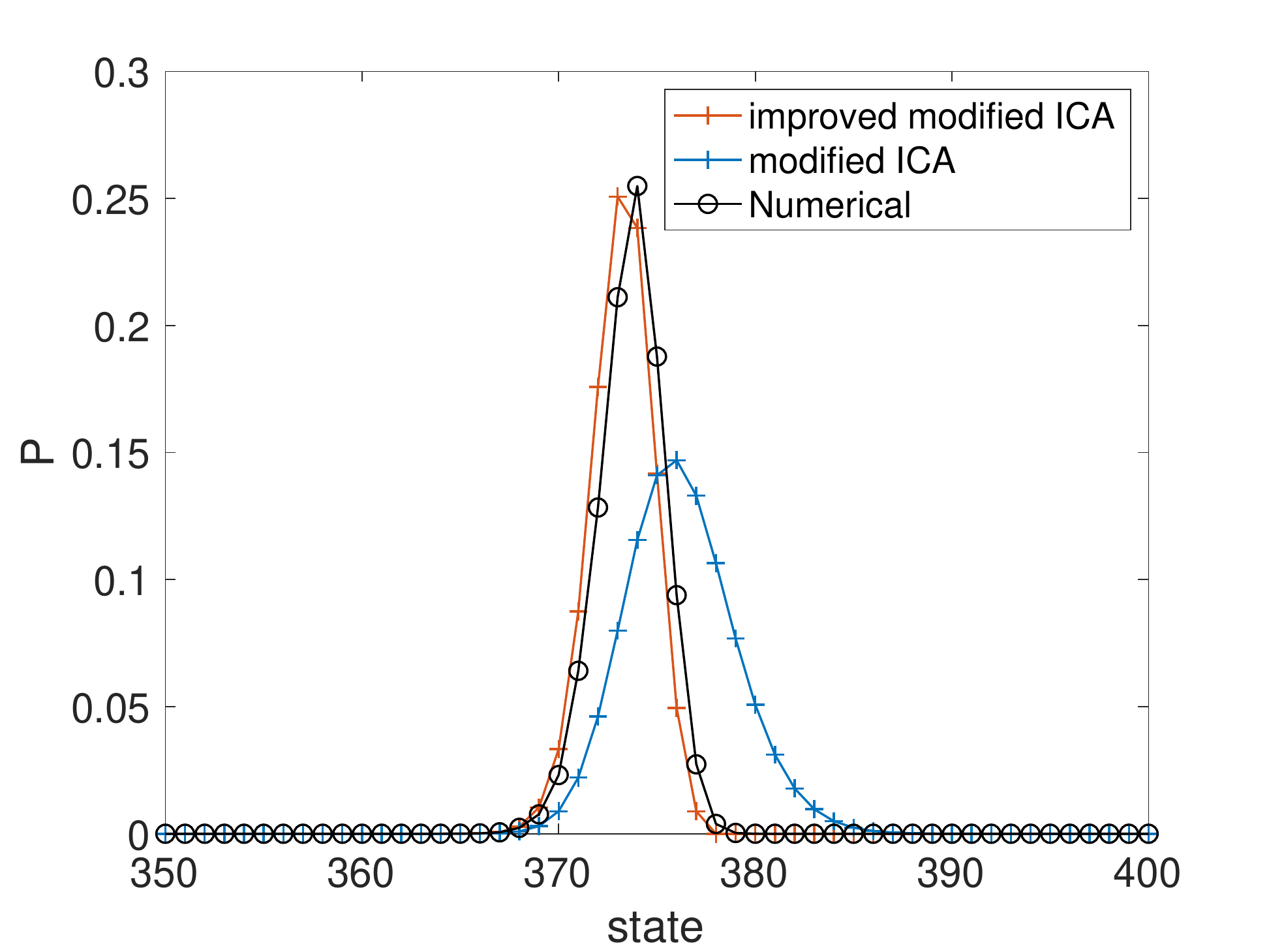}
\caption{Comparison of the modified ICA (blue crosses) and an improved modified ICA (red crosses, see text) with a numerical solution of the Schr\"odinger equation (black open circles). We show the probability to be in an adiabatic eigenstate at the end of the forward sweep where the initial state is the ground state, so that there are no effects of path interference. The system parameters are $N=1000$, $2\Omega T=10000$, $\Delta_0/\Omega=-\Delta_I/\Omega=2$. Lines between data points have been added as guides to the eye.}
\label{fig:test_ICA}
\end{figure}

\begin{figure}
\includegraphics[width=.45\textwidth]{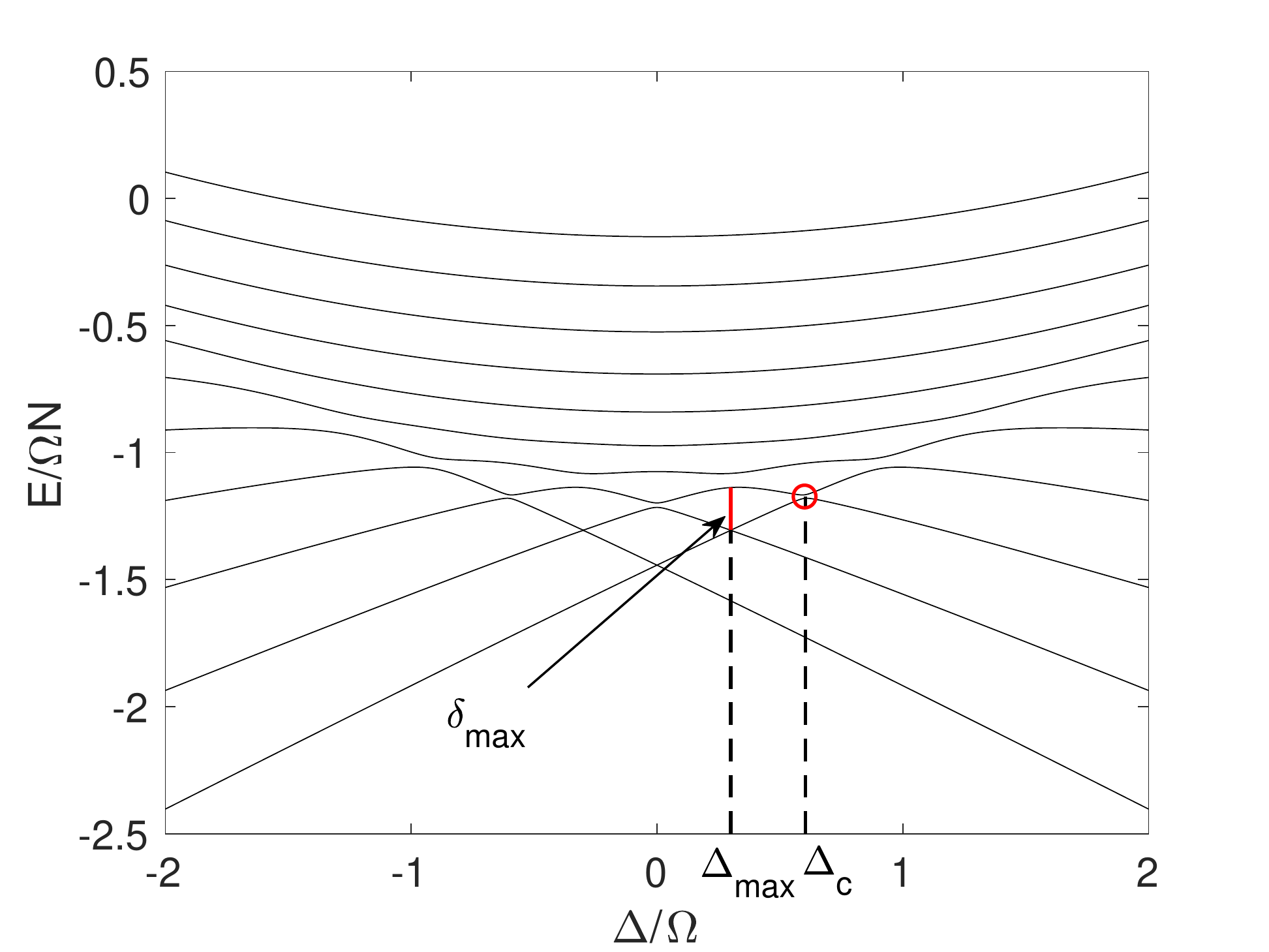}
\caption{Definition of $\delta_{\mathrm{max}}$, $\Delta_{\mathrm{max}}$ and $\Delta_c$ for the avoided crossing indicated by the red circle.}
\label{fig:delta_max}
\end{figure}

We can now test the accuracy of the modified ICA in our system. To do so we consider the large particle number $N=1000$ and total sweep time $2\Omega T=10000$ for a sweep from $\Delta_I/\Omega=-2$ to $\Delta_0/\Omega=2$, as in Fig.~\ref{fig:N=1000}. In general there will be many different paths along the adiabatic levels, branching at avoided crossings, to connect a given initial and final state. Therefore, if we describe the system state in the adiabatic basis, the interference between the amplitudes of different paths has to be taken into account; yet typically this interference is not included in the ICA. If the initial state is chosen to be the ground state, however, then there is no interference during the forward sweep because there exists only a single path connecting the initial ground state at $\Delta_I$ and any given state at $\Delta_0$ (see Fig.~\ref{fig:spectrum}), so that the dynamical phase that is responsible for interference plays no role. We can  therefore use this setup to test the modified ICA. Fig.~\ref{fig:test_ICA} shows the probability distribution over adiabatic eigenstates at the end of the forward sweep ($t=0$, $\Delta(t)=\Delta_0$), obtained from a direct numerical solution of the Schr\"odinger equation and from the modified ICA.
We find that the modified ICA gives reasonably good results given the fact that there are 1001 states. The modified ICA can be greatly improved, though, if we do not use the difference of the slopes of the diabatic levels in the Landau-Zener formula, but rather the difference of the asymptotic slopes of the adiabatic levels. To find this difference of the asymptotic slopes for a given avoided crossing of the levels $i$ and $j$ at $\Delta=\Delta_c$, we search for the nearest local maximum of the level separation $\delta_{ij}^{\mathrm{max}}$ of the involved adiabatic levels at $\Delta=\Delta_{\mathrm{max}}$ and set the slope to $\alpha_{ij}=\delta_{ij}^{\mathrm{max}}/|\Delta_{\mathrm{max}}-\Delta_c|$, see Fig.~\ref{fig:delta_max}. 
Note that this construction does not work for the highest energy levels, far away from the swallowtail, as here there are no longer any clear maxima and minima of the level separation. This does not pose a problem, though, because for our slow sweep the diabatic transition probability is essentially zero at these high energies. Moreover, as we have seen in Figs.~\ref{fig:N=30}--\ref{fig:N=1000}, the adiabatic levels for which the level separation has no clear local extrema play no role in the evolution. If we label the energy gap at the avoided crossing by $\delta_{ij}$, the probability for a diabatic transition from the adiabatic level $i$ to $j$ is given by
\begin{equation}
P_{ij}=e^{- \frac{\pi \delta_{ij}^2}{2 \dot \Delta \alpha_{ij}}},
\end{equation}
as stated in the main text. Note that in general there are multiple avoided crossings between two adiabatic levels $i$ and $j$, so that all the quantities mentioned above should have an additional index labelling the different avoided crossings. Note also that the first avoided crossing (between the first and second adiabatic level) has to be treated differently, because also for it there is no local maximum of the corresponding level separation, but that it is only this first pair of levels which need this special treatment, and the special treatment is simple. If we look at Fig.~\ref{fig:spectrum} then it is clear that we can just use the level separation at the value of $\Delta$ where second and third adiabatic levels have an avoided crossing as value for $\delta_{12}^{\mathrm{max}}$. 

In effect this ``improved modified'' procedure gives a much better local two-level approximation of each avoided crossing than the ordinary modified ICA; accordingly this improved modified ICA (red crosses in Fig.~\ref{fig:test_ICA}) gives a much better approximation to the true probability distribution after the sweep. In this paper we have used only this improved version of the modified ICA, although for brevity we have referred to it simply as ``the Landau-Zener approximation''. The drawback of this method compared to the usual ICA is that the Hamiltonian has to be diagonalized at a large number of $\Delta$ values to find the minima and maxima of the separation of the adiabatic levels, whereas the slopes and coupling matrix elements in the usual ICA can be read off the Hamiltonian. Even for $N=1000$, however, this can be done on a desktop computer within just a few days.

\end{document}